\documentclass[12pt,a4paper]{article}
\usepackage[margin=1 in]{geometry}
\usepackage{hyperref}
\hypersetup{
  colorlinks   = true,    
  urlcolor     = black,    
  linkcolor    = RoyalBlue,    
  citecolor    = OliveGreen      
}
\usepackage{graphicx}

\usepackage{url,lineno,microtype,subcaption}
\usepackage{amsmath,amsthm,amssymb,stmaryrd} 
\usepackage{pifont,textcomp} 
\usepackage[dvipsnames]{xcolor}
\usepackage{multirow}
\usepackage{epstopdf}
\usepackage[onehalfspacing]{setspace}
\newtheorem{prop}{Proposition}
\usepackage{enumitem}
\newcommand\blfootnote[1]{%
  \begingroup
  \renewcommand\thefootnote{}\footnote{#1}%
  \addtocounter{footnote}{-1}%
  \endgroup
}
\begin{document}
\hfill 
  \begin{center}
    {  \Large \textsf Model-based assessment of the Role of Uneven Partitioning of Molecular Content on Heterogeneity and Regulation of Differentiation in CD8 T-cell Immune Responses }
        \end{center}  
        
  \begin{center}
     {Simon Girel\,$^{1,2,*}$, Christophe Arpin\,$^{3}$, Jacqueline Marvel\,$^{3}$, Olivier Gandrillon\,$^{2,4}$ and Fabien Crauste\,$^{1,2,*}$}
     \vspace{0.6cm}
     
  \textit{$^{1}$Univ Lyon, Universit\'{e} Claude Bernard Lyon 1, CNRS UMR 5208, Institut Camille Jordan, 43 blvd. du 11 novembre 1918, F-69622 Villeurbanne cedex, France \\
$^{2}$Inria, Villeurbanne, France \\
$^{3}$Univ Lyon, ENS de Lyon, Univ Claude Bernard, CNRS UMR 5308, INSERM
U1111, Centre International de Recherche en Infectiologie, Universit\'{e} Lyon 1, Lyon
F-69007, France\\
$^{4}$Univ Lyon, ENS de Lyon, Univ Claude Bernard, CNRS UMR 5239, INSERM U1210,
Laboratory of Biology and Modelling of the Cell, 46 all\'{e}e d’Italie Site Jacques Monod, F-69007 Lyon, France  \\
$^{*}$\textsf{girel@math.univ-lyon1.fr; fabien.crauste@u-bordeaux.fr }}
        \end{center}

\blfootnote{Preprint submitted to \textit{Frontiers in Immunology (2018)}. Final version accepted on 01/2019,  doi : 10.3389/fimmu.2019.00230.} 

 \hrulefill 
\vspace{1 cm} 
\begin{abstract}

Activation of naive CD8 T-cells can lead to the generation of multiple effector and memory subsets. Multiple parameters associated with activation conditions are involved in generating this diversity that is associated with heterogeneous molecular contents of activated cells.  Although naive cell polarisation upon antigenic stimulation and the resulting asymmetric division  are known to be a major source of heterogeneity and cell fate regulation, the consequences of stochastic uneven partitioning of molecular content upon subsequent divisions  remain unclear yet. Here we aim at studying the impact of uneven partitioning on molecular-content heterogeneity and then on the immune response dynamics at the cellular level. To do so, we introduce a multiscale mathematical model of the CD8 T-cell immune response in the lymph node. In the model, cells are described as agents evolving and interacting in a 2D environment while a set of differential equations, embedded in each cell, models the regulation of intra and extracellular proteins involved in cell differentiation. Based on the analysis of \textit{in silico} data at the single cell level, we show that immune response dynamics can be explained by the molecular-content heterogeneity generated by uneven partitioning at cell division. In particular, uneven partitioning acts as a regulator of cell differentiation and induces the emergence of two coexisting sub-populations of cells exhibiting antagonistic fates. We show that the degree of unevenness of molecular partitioning, along all cell divisions, affects the outcome of the immune response and can promote the generation of memory cells.

\textbf{Keywords: }  Multiscale modeling, immune response, asymmetric division, agent-based models, immune memory 
\end{abstract}

\section{Introduction}

Following acute infection, the activation of naive CD8 T-cells by antigen presenting cells (APCs) triggers the synthesis of proteins controlling cell proliferation and differentiation up to the memory state. While CD8 T-cell population dynamics have been widely described, it is of great interest to better understand the molecular mechanisms driving the CD8 T-cell response. In particular, determining the effects of molecular events on the generation of memory cells is necessary for vaccine design improvement.  \textit{In vivo} and \textit{in vitro} studies have demonstrated that a single presentation of the antigen to naive CD8 T-cells is sufficient to trigger a complete CD8 T-cell immune response \cite{Stipdonk,Kaech2001,Wherry2004,Badovinac2002,Wong2001}. Then, once initiated, antigen-independent molecular pathways drive a program of CD8 T-cell proliferation and differentiation \cite{ANTIA2003,Blueprint}. 

 The CD8 T-cell immune response occurs through four main phases. First the activation of naive CD8 T-cells in secondary lymphoid organs such as lymph nodes (LN) or spleen by APCs through MHC class I antigenic peptide/T-cell receptor (TCR) binding, surface co-receptor/ligands interactions and soluble cytokines secretion. Once activated, CD8 T-cells proliferate quickly during the expansion phase, which expands the initial population by a factor of $10^3$ to $10^5$ \cite{ANTIA2003,Blattman2002}.  Concomitantly, activated cells differentiate into effector cells, able to kill infected cells through cytotoxicity. At the end of the expansion phase, known as the peak of the response, the CD8 T-cell population begins a contraction phase, where most of the responding cells die yet leaving a quiescent population of cells with strong re-activation potential: the memory cells. The memory cell population survives the contraction phase and may remain for years in the organism (memory phase) to ensure faster and stronger host-protection against subsequent infection by the same pathogen. 

The responding effector population is composite and two subsets with antagonistic fates have been described \cite{Joshi2007}: memory precursor effector cells (MPEC)  and short-lived effector cells (SLEC), characterised by the expression of two proteins KLRG1 and CD127 (IL-7 receptor). Both MPEC (KLRG1$^{lo}$CD127$^{hi}$) and SLEC (KLRG1$^{hi}$CD127$^{lo}$) express effector features (cytotoxicity, proliferation) but MPEC  are capable of differentiation into memory cells while SLEC are destined to die during the contraction phase \cite{Joshi2007}. Thus CD8 T-cell population dynamics arise from cell phenotypic heterogeneity, itself resulting from molecular-content heterogeneity.

Among the genes, transcription factors and proteins involved in the CD8 T-cell response, some seem to play key roles in the differentiation processes.
Transcription factors Tbet and Eomesodermin (Eomes) appear to play critical roles in the acquisition of effector and memory phenotypes. It has been shown that the expression of Tbet induces the development of SLEC and represses the development of MPEC profiles \cite{Joshi2007,Huang2015,Lazarevic2013}. 
 Eomes is not involved in the SLEC versus MPEC fate choice \cite{Banerjee,Kaech2012}. However, Eomes is necessary for the development of several properties of memory cells (survival, lymph node homing capacities, responsiveness to second infection \cite{Lazarevic2013,Banerjee,Munitic2010}). Along the differentiation from effector to memory, the concentration of Tbet in a CD8 T-cell decreases, while the concentration of Eomes increases \cite{Lazarevic2013,Joshi2011}. 
 
 Since a unique initial antigenic signal can trigger a complete response, additional mechanisms are necessary to generate the observed molecular-content heterogeneity.
 Chang \textit{et al.} \cite{Arsenio2015,Chang2007,Chang2011,Ciocca2012} showed that TCR binding to MHC-class-I peptide-complex results in polarised segregation of proteins in activated  CD8 T-cell: some proteins migrate on the TCR side of the T-cell, other migrate on the opposite side.  The subsequent division of the activated CD8 T-cell splits the mother cell perpendicularly to the polarisation axis, such that  the daughter cell coming from the TCR side (proximal cell) receives more proteins associated to effector lineage, including Tbet, while the  other one (distal cell)  receives more proteins associated to memory lineage. Asymmetric division of polarised naive CD8 T-cells appears to be one of the major mechanisms regulating CD8 T-cell fate decision.
 
 Nevertheless, the role of asymmetric division of polarised naive cells in the T-cell differentiation process appears to be controversial  \cite{Cobbold2018}. While there are several evidences for asymmetric division of polarised naive CD8 T-cells \cite{Pham2014}, it remains uncertain how this polarisation quantitatively depends on the affinity of the TCR for the MHC-class-I peptide-complex, the duration of the binding,  external chemokines and interactions with homotypic CD8 T-cells \cite{Pham2014}.  Since  the  asymmetric partitioning of Tbet  has been evidenced in mice CD8 T-cells, it will be considered hereafter.

Less is known about the partitioning of molecular content in the course of subsequent cell divisions. However, several studies support the hypothesis that when a cell divides, a random, uneven partitioning of the molecular content occurs \cite{Block1990,Bocharov2013,Golding2005,Huh2010,Luzyanina2013,Sennerstam88,Thomas2018,Kinkhabwala2014}.  Partitioning of CFSE dye, a cell staining dye used to track cell proliferation through dye dilution, during lymphocyte proliferation has been mathematically studied by Luzyanina \textit{et al.} \cite{Bocharov2013,Luzyanina2013}. Based on comparison with  \textit{in vitro} experimental data,  these studies suggest that uneven partitioning, which does not result from cell polarisation, occurs at T-cell division.

We emphasize that the  \emph{asymmetric} first division of naive cells, which goes through an active polarisation of the cell, has to be distinguished from the  random partitioning of the molecular content during the subsequent divisions of  non-polarised cells,  hereafter referred to as \textit{uneven} partitioning \cite{Kinkhabwala2014}.

In a recent work (Girel and Crauste \cite{Girel2018}), we studied how stochastic uneven molecular partitioning, repeated at each cell division, could regulate the effector versus memory cell-fate decision in a CD8 T-cell lineage. To do so, we analysed  an impulsive differential equation describing the concentration of the protein Tbet in a CD8 T-cell subject to divisions, where impulses were associated with uneven partitioning of Tbet. In this work, high and low Tbet concentrations were associated with effector and memory phenotypes, respectively. We concluded that, for a low degree of unevenness of molecular partitioning, a CD8 T-cell expressing a moderate concentration of Tbet can still generate both memory and effector cells. If the concentration of Tbet in this cell is high or low enough, the phenotype of the cell and its progeny becomes irreversible, with low Tbet-expresser  and high Tbet-expresser differentiating in memory or effector cells,  respectively. Moreover, our study indicates that the increase in cell cycle length throughout the immune response \cite{Kinjyo2015,Yoon2010} favours irreversible cell differentiation. 

Several works (see \cite{Eftimie2016} and the references therein), focused on modeling  molecular mechanisms of the immune response coupled to cell population dynamics. Most of these works involve agent-based models. 

Gong \textit{et al.} \cite{Gong2013,Gong2014} developed a two-compartment model to study how the number of dentritic cells and the level of MHC-peptides on their membrane influence the size and composition of T-cell populations. Since they did not model any dynamics at the molecular level, they were limited in studying the molecular origins of cell differentiation and  heterogeneity.


Prokopiou \textit{et al.} \cite{Prokopiou} and Gao \textit{et al.} \cite{Gao2016} designed a multi-scale agent-based model of the early CD8 T-cell immune response (Day 3 to 5.5 post-infection).  At the population scale, a discrete population of CD8 T-cells and APCs in a LN is modeled by a cellular Potts model (CPM) \cite{GGH}. At the molecular scale, the dynamics of a simplified molecular regulatory network (MRN) containing some key molecular factors is modeled by a system of differential equations, embedded in each cell of the population, whose state determines cell phenotype and fate. Cells communicate with each other through cell-cell contact and secretion of the cytokine IL2 such that the environment of a cell affects its molecular profile. Parameter calibration resulted in good agreement with \textit{in vivo} data of an immune response in murine LN after influenza infection, at both cellular and molecular levels. 

The model presented in this article has been developed from the multi-scale agent-based model previously introduced in \cite{Prokopiou,Gao2016}. Since the authors in \cite{Prokopiou,Gao2016} focused on early events following CD8 T-cell activation, they did not consider processes leading to the generation of memory cells. We enriched their model in order to study a complete response, from the activation of naive cells to the generation of memory cells. In particular, Eomes has been added to the MRN. 

In this paper, we are interested in understanding how, from the activation of naive CD8 T-cells, an antigen-independent regulation of intra-cellular molecular content can drive a complete CD8 T-cell response. We particularly focus on the role of molecular-content heterogeneity among a CD8 T-cell population in the generation of memory cells. We first verify our model's ability to reproduce \textit{in vivo} data at both cellular and molecular scales. Then we study,  in an \textit{in silico} CD8 T-cell population, the impact of molecular-content heterogeneity on the emergence of sub-populations, characterised by their expression of proteins Tbet and Eomes. We discuss how uneven distribution of molecular content at cell division affects the cellular dynamics (population size, cell differentiation and death) and suggest that memory cell generation  efficiency is maximal for a moderate degree of unevenness. Finally, we show that memory cells generated by our model are able to reproduce some features of a secondary CD8 T-cell immune response. Indeed, when restimulated by antigen \textit{in silico} they generate more  cells at the peak of the response and in the memory phase. 

\section{Material, Methods and Model}
\label{SectionModel}
\subsection{Data}
$4\times 10^5$ naive CD8 T-cells from CD45.1+ F5 TCR transgenic mice (B6.SJL-Ptprc$^a$Pepc$^b$/BoyCrl-Tg(CD2-TcraF5, CD2-TcrbF5)1Kio/Jmar) recognizing the
NP68 epitope were transferred intravenously in
congenic CD45.2+ C57BL/6 mice (C57BL6/J). The day after recipient mice were inoculated intranasally with $2\times10^5$ PFU (plaque forming units) of a vaccinia virus expressing the NP68 epitope \cite{Jubin2012}. From day 4 to day 22 post-infection,
the spleens of infected animals where harvested and the number of F5 transgenic CD8 responder T-cells was assessed by flow cytometry, based on CD8/CD45.1/CD45.2 expression, to distinguish F5 TCR-transgenic responder (CD45.1$^+$CD45.2$^-$) from host (CD45.1$^-$CD45.2$^+$) CD8 T-cells. Naive (CD44$^-$ Mki67$^-$ Bcl2$^+$), effector (CD44$^+$ Bcl2-) and memory (CD44$^+$ Mki67$^-$ Bcl2$^+$) CD8 T-cells have been identified \cite{Crauste2017}. All experimental procedures were approved by an animal experimentation ethics committee (CECCAPP; Lyon, France), and accreditations have been obtained
from the French government.

OT1 CD8 T cells mRNA expression data time courses come from the ImmGen project  (\url{http://www.immgen.org}). According to the information provided on ImmGen.org, the \textit{in vivo} mRNA data
(Figure \ref{ResultsMolec}) were generated for OT1 T-cells stimulated in similar
experimental settings \textit{i.e.} the response of transferred OT1 TCR-transgenic CD8 T-cells following infection by vesicular stomatitis virus expressing their cognate antigen.

\subsection{Molecular regulation and IL2 diffusion}
\label{SectionReseau}
\begin{figure}[t!]
\centering
\includegraphics[width = 14 cm]{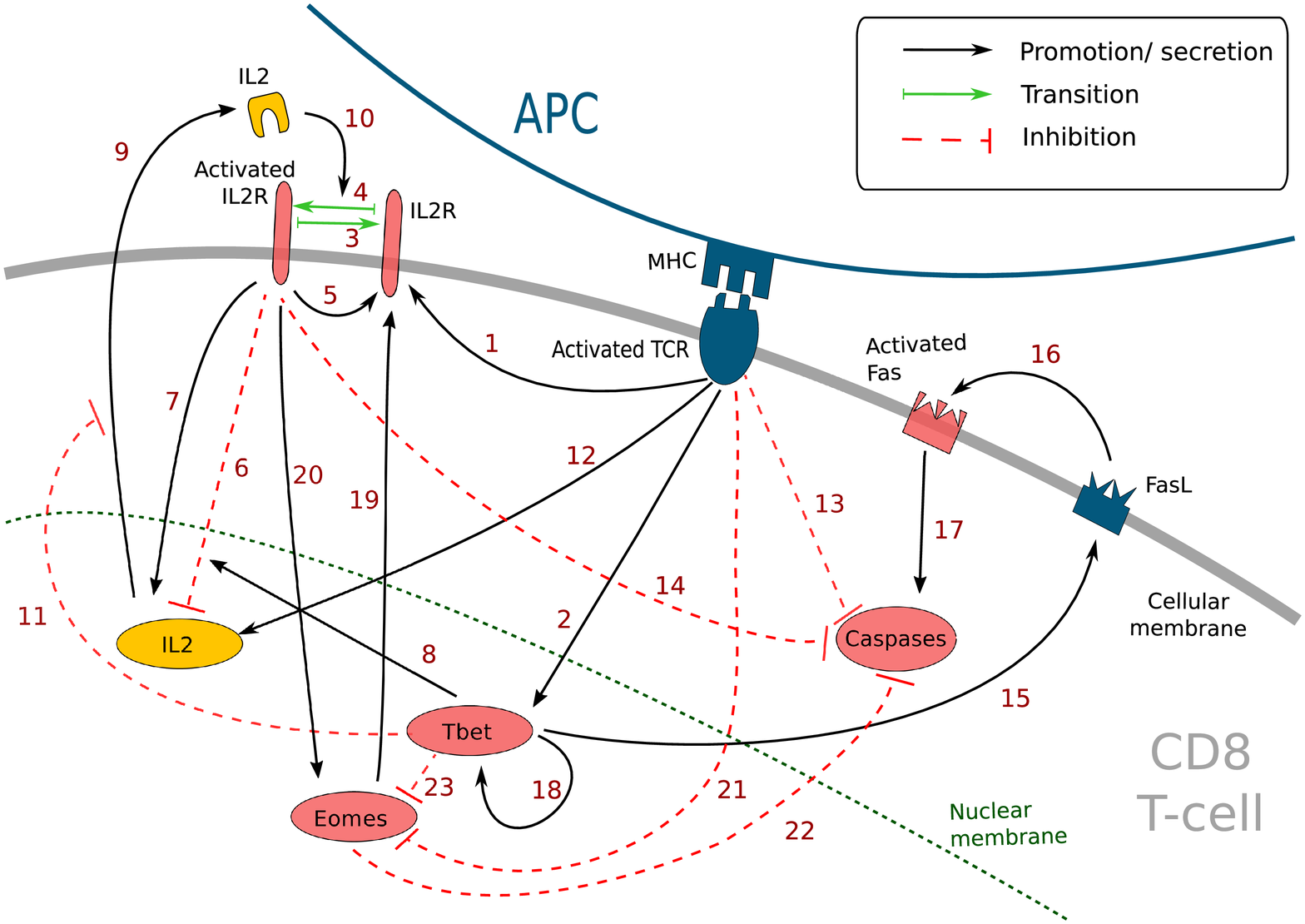}
\caption{Simplified molecular regulatory network in a CD8 T-cell. Red molecular factors dynamics are described by equations (\ref{EqR})-(\ref{EqE}); yellow molecular factors dynamics are described by equation (\ref{SecretionIL2});  black arrows: promotion or secretion; green arrows: transition between activated and non-activated form of IL2R; red dashed arrows: inhibition. The meaning of the numbered arrows is reported in Table \ref{legendeReseau}.}
\label{ModeleBio}
\end{figure}

We aim at describing the molecular regulation within each CD8 T-cell during a response to an acute infection, and how the dynamical molecular state of a cell characterises its differentiation stage.
We present on Figure \ref{ModeleBio} the MRN that will be used throughout this manuscript and give a detailed description in Table~\ref{legendeReseau}. It contains several key molecular factors involved in CD8 T cell proliferation, differentiation, apoptosis, and cell communication. This is an updated version of the MRN developed in \cite{Prokopiou,Gao2016} that was limited to the description of differentiation up to the effector stage. To account for differentiation into memory cells, we introduced the protein Eomes and its interactions with the rest of the network as documented in the literature. Indeed, Eomes is involved in the development of essential properties of memory cells such as survival, lymph node homing capacities or faster response to antigenic stimulation \cite{Lazarevic2013,Banerjee,Munitic2010}.

\subsubsection{Molecular regulatory network}
This MRN is initiated upon antigen presentation to a naive CD8 T-cell, through the engagement of the TCR. Antigenic stimulation triggers the synthesis of interleukine-2 (IL2) by the CD8 T-cell  
 and the production of IL2 receptors (IL2R) on the cell membrane \cite{Hoyer2008}. The synthesised IL2 is then released in the environment and can bind its receptor \cite{Feinerman2010} to form IL2-IL2R complex, hereafter referred to as activated IL2R.    
Activated IL2 receptors enhance the expression of IL2 receptors \cite{Hoyer2008} as well as IL2 synthesis  \cite{Hoyer2008}. In the meantime, activated IL2 receptors, jointly with protein Tbet (see below), inhibit the activation of the \textit{IL2} gene through the action of the mediator protein Blimp1 \cite{Martins2008,Yeo2010}.

Antigenic stimulation independently stimulates Tbet synthesis \cite{McLane2013}, a protein involved in the acquisition of cell cytotoxicity. 
Indeed, Tbet is known to induce the expression of Fas ligand (FasL)  \cite{Sullivan2003}, 
a transmembrane protein that can bind to the transmembrane protein Fas to induce cell apoptosis via the activation of Caspases in the Fas-expressing cell \cite{Bouillet2009}.  Caspases are a family of proteins playing essential role in cell apoptosis \cite{Strasser2009}. There exist several types of Caspases involved in CD8 T-cell apoptosis yet, for the sake of simplicity, we aggregated them in a unique variable $[Cas]$. 
 Moreover, Tbet induces its own synthesis (via the gene \textit{Tbx21}) \cite{Kanhere2012,Shin2009}.

Eomes expression, involved in the acquisition of memory phenotype \cite{Banerjee}, is first inhibited during the activation phase due to engagement of the TCR (via activation of the Akt/mTOR pathway and inhibition of FOXO1 and TCF7)  \cite{Blueprint,Kaech2012,Suresh}. Eomes is induced later \cite{Lazarevic2013,Cruz} and its expression is enhanced by the activation of IL2 receptors \cite{Blueprint,Kaech2012,Li2013}. 
Eomes promotes the development of new IL2 receptors on cell membrane  \cite{Munitic2010}.

The activation of IL2 receptors, of the TCR and the protein Eomes prevents apoptosis by inhibiting the activation of Caspases, in particular through the mediator protein Bcl2 \cite{Banerjee,Ewings2007,Kelly2003}

{\renewcommand{\arraystretch}{1.2}
\begin{table}[h!]
\small
\begin{tabular}{|clc|}
\hline 
n° & Description & R\'{e}f. \\ \hline
1 & Activated TCR induces the development of IL2 receptors & \cite{Feinerman2010,Boyman2010}  \\ 
2 & Activated TCR induces the synthesis of Tbet & \cite{McLane2013}  \\ 
3 & Deactivation of activated IL2 receptors  &  \cite{Feinerman2010} \\ 
4 & Activation of IL2 receptors & \cite{Feinerman2010}  \\ 
5 & Activated IL2 receptors induce the development of new IL2 receptors &  \cite{Hoyer2008} \\ 
6 & Activated IL2 receptors inhibit the expression of the \textit{IL2} gene (via Blimp1) &  \cite{Martins2008} \\ 
7 & Activated IL2 receptors induce the expression the \textit{IL2} gene & \cite{Hoyer2008}  \\ 
8 & Tbet enhances the inhibition of  the \textit{IL2} gene by activated IL2 receptors &  \cite{Yeo2010} \\ 
9 & Internal IL2 is secreted in extracellular environment &   \cite{Feinerman2010} \\ 
10 & External IL2 binds the non-activated IL2 receptors to activates them &  \cite{Hoyer2008} \\ 
11 & Tbet inhibits the secretion of IL2 & \cite{last,Hwang2005,Szabo2000}  \\ 
12 & TCR activation activates \textit{IL2} gene (via Erk) & \cite{Hoyer2008}  \\ 
13 & TCR activation inhibits the activation of Caspases (via Erk, Bim, Bax and Bcl2)  & \cite{Ewings2007}  \\ 
14 & Activated IL2 receptors inhibit the activation of Caspases (via Stat5, BAX et Bcl2) &  \cite{Kelly2003} \\ 
15 & Tbet induces the expression of FasL & \cite{Sullivan2003}  \\ 
16 & FasL activates Fas through cell contact &  \cite{Bouillet2009}  \\ 
17 & Activated Fas induces Caspases activation &  \cite{Bouillet2009} \\ 
18 & Tbet activates \textit{Tbx21} and induces the synthesis of Tbet (positive feedback loop) &  \cite{Kanhere2012,Shin2009}  \\ 
19 & Eomes induces the expression of IL2 receptors & \cite{Munitic2010}  \\ 
20 & Activated IL2 receptors induce the expression of \textit{Eomes} (via Runx3) & \cite{Blueprint,Kaech2012,Li2013}  \\ 
21 &  Activated TCR inhibits \textit{Eomes} gene expression (via Akt, mTOR, Tcf1, Foxo1) & \cite{Blueprint,Kaech2012,Suresh}   \\ 
22 & Eomes inhibits the activation of Caspases (via Bcl2) & \cite{Banerjee}  \\ 
23 & Tbet inhibits the expression of \textit{Eomes} (via IFN$\gamma$, IL12R) & \cite{Afkarian2002,Baumjohann2013}  \\ 
\hline 
\end{tabular} 
\captionof{table}{Description of the molecular signalling pathways in Figure \ref{ModeleBio} and corresponding bibliographic references.}
\label{legendeReseau}
\end{table} }

\subsubsection{Intracellular molecular dynamics}\label{SectionIntra}
Based on the above-described reactions, and from the equations used in \cite{Prokopiou,Gao2016},  we describe the dynamics of the concentrations of non-activated IL2 receptors ($[R]$), activated IL2 receptors ($[L\bullet R]$), Tbet ($[Tb]$), activated Fas ($[Fs^*]$), Caspases ($[Cas]$) and Eomes ($[E]$) in a CD8 T-cell with the following system of equations

\begin{flalign}
  \dfrac{\mathbf{d}}{\mathbf{d}t} [R] =& \lambda_{R1}f_{APC}+(\mu_{IL2}^-+\lambda_{R2})[L\bullet R]+\lambda_{E1}[E] 
 -\left(\mu_{IL2}^+[IL2^{cm}]+k_R\right)[R], \label{EqR}\\
 \dfrac{\mathbf{d}}{\mathbf{d}t} [L\bullet R] =&  \mu_{IL2}^+[IL2^{cm}][R]-\mu_{IL2}^-[L\bullet R]-k_e[L\bullet R],\label{EqLR} \\
 \dfrac{\mathbf{d}}{\mathbf{d}t} [Tb]=&   \lambda_{T1}f_{APC}+\lambda_{T2}\dfrac{[Tb]^n}{\lambda^n_{T3}+[Tb]^n}-k_T[Tb], \label{EqTb}\\
 \dfrac{\mathbf{d}}{\mathbf{d}t} [Fs^*]=&  H\mu_F^+[Tb^{cm}]\left(\dfrac{\lambda_F}{k_{F}}-[Fs^*]\right)-\mu_F^-[Fs^*]-k_{F}[Fs^*],\label{EqFs} \\
 \dfrac{\mathbf{d}}{\mathbf{d}t}[Cas] =&  G\lambda_{c1}\dfrac{1}{1+\lambda_{c2}[L\bullet R]}\cdot \dfrac{1}{1+\lambda_{c3}f_{APC}}\cdot \dfrac{1}{1+\lambda_{E2} [E]}  +\lambda_{c4}[Fs^*]-k_c[Cas],\label{EqCas}\\
  \dfrac{\mathbf{d}}{\mathbf{d}t}[E] =&  \dfrac{1}{1+\lambda_{E5}f_{APC}}\cdot \left( \dfrac{\lambda_{E3}[L \bullet R]}{\lambda_{E6}+[L\bullet R]}+\dfrac{G\lambda_{E4}}{1+\lambda_{E7}[Tb]}\right)-k_E[E].\label{EqE}
\end{flalign}

All parameters are positive. Parameters $\lambda$ are associated to induction and inhibition effects, $\mu$ are associated to activation and deactivation of transmembrane proteins and $k$ are degradation and dilution rates. The concentrations of System (\ref{EqR})-(\ref{EqE}) are assumed to be null in naive CD8 T-cells, and remain null until TCR engagement.

The effects of the external environment on the intracellular system (\ref{EqR})-(\ref{EqE}) are taken into account through five variables.  The variable $f_{APC}$ (equations (\ref{EqR}), (\ref{EqTb}), (\ref{EqCas}) and (\ref{EqE})) is equal to the number of APCs bound to the considered CD8 T-cell and accounts for TCR engagement. The variable $G$ (equations (\ref{EqCas}) and (\ref{EqE})) is equal to $0$ in naïve CD8 T-cells and to $1$ otherwise, \textit{i.e.} in cells that have already met with an APC. It accounts for the fact that up-regulation of Caspases and Eomes described by parameters $\lambda_{c1}$ and $\lambda_{E4}$ is not active in naive cells. The variable $H$  (equation (\ref{EqFs})) accounts for the expression of FasL by effector and memory T-cells and for the activation of Fas through cell contact. Hence, $H$ is equal to 1 in a non-naive considered CD8 T-cell in contact with an effector or a memory CD8 T-cell, and equal to 0 otherwise. The variable $[IL2^{cm}]$ is equal to the concentration of IL2 at the cell membrane, in the extracellular environment. Finally, $[Tb^{cm}]$ is defined as the sum of Tbet concentrations in effector and memory CD8 T-cells in contact with the considered CD8 T-cell and acts as a proxy for the expression of Fas in those cells.

We introduced the variable $[E]$ and the associated equation (\ref{EqE}) to the system used in \cite{Gao2016} in order to account for  the synthesis of protein Eomes and its interactions with other molecular factors.
The term $\lambda_{E1}[E]$ in (\ref{EqR}) accounts for the up-regulation of IL2 receptors by Eomes. Eomes also limits cell apoptosis by activating \textit{Bcl-2} gene, as do IL2 and activated TCR. This communal target explains the multiplicative form of the inhibition of Caspases by Eomes, IL2 and TCR in equation (\ref{EqCas}). We also introduced the function $G$ in (\ref{EqCas}) to update the dynamics of Caspases concentration from  \cite{Prokopiou,Gao2016}.

 The positive feedback loop on Tbet is modeled with an order $n$ Hill function in order to allow bistable behaviour of Tbet. As discussed in the introduction, the concentration of protein Tbet can be associated to the level of differentiation of an effector CD8 T-cell, with high level of Tbet correlating with fully differentiated effector cell, while low Tbet levels are associated to memory precursor effector cells.  Proposition \ref{PropTbet} below, reproduced from \cite{Girel2018}, gives necessary and sufficient conditions to allow bistable behaviour of Tbet concentration.

\begin{prop}[\cite{Girel2018}]
\label{PropTbet}
 Assume $f_{APC}=0$, $n>1$ and $\lambda_{T2}(n-1)^{\frac{n-1}{n}}>nk_T \lambda_{T3}$, then equation (\ref{EqTb}) has exactly three non-negative steady states: $0<[Tb]_u<[Tb]_s$, such that $0$ and $[Tb]_s$ are locally asymptotically stable and $[Tb]_u$ is unstable.
\end{prop}
In the following, we will assume that the conditions $n>1$ and $\lambda_{T2}(n-1)^{\frac{n-1}{n}}>nk_T \lambda_{T3}$ are fulfilled (see Section \ref{SectionSLECMPEC}).


System (\ref{EqR})-(\ref{EqE}) is embedded in every CD8 T-cell. Nevertheless, cell-cell contacts
, stochastic events (cell cycle length, protein distribution at division) and external concentrations of IL2 
affect the evolution of the system such that each CD8 T-cell develops a unique molecular profile based on its own history.

\subsubsection{Extracellular IL2 diffusion} \label{SectionIL2}
The secretion of IL2 by CD8 T-cells and its isotropic diffusion  in the extracellular domain (with periodic boundary conditions) are modeled by the following PDE, introduced by Prokopiou \textit{et al.} \cite{Prokopiou},
\begin{equation}
\dfrac{\partial [IL2]}{\partial t}=D\nabla ^2[IL2]+\left(\lambda_{R3}\dfrac{[L\bullet R]}{\lambda_{R4}+[L \bullet R]}+\lambda_1f_{APC}\right)\dfrac{1}{1+\lambda_{T4}[Tb]}-\delta[IL2], \label{SecretionIL2}
\end{equation}
where $[IL2]$ is the IL2 concentration. CD8 T-cells react to extracellular IL2 through their IL2 receptors by means of the $[IL2^{cm}]$ term, in (\ref{EqR})-(\ref{EqLR}), defined as the sum of $[IL2]$ at the considered cell membrane.

\subsection{Cell differentiation and division}
\label{regles du modele}
Rules controlling cell division (including protein distribution at the division), apoptosis and differentiation are summarised in Table \ref{Proprietes} and detailed hereafter. It must be noted that cells properties  result from their molecular profile. For example, the properties observed \textit{in vivo} in memory cells (survival, low IL2 secretion, low cytotoxicity) are not imposed by model rules but acquired as a consequence of their molecular profile. One exception is cell cycle duration (see \ref{SectionCycle}). 
{\renewcommand{\arraystretch}{1.3}
\begin{table}\begin{center} 
\begin{tabular}{|c ||c |c| c |c |c| c|}
\hline
\multirow{2}{*} \ \ \ \ \ \ \ \ \ \  \ $\backslash$ Property & Division & Apoptosis & IL2 & FAS & FasL \\Cell type $\backslash$\ \ \ \ \  \ \ \ \ & & & secretion & expression & expression \\ 
\hline     
APC  &  \ding{109} & \ding{52} & \ding{109} & \ding{109} & \ding{109}   \\ 
\hline 
Naive  &  \ding{109} &  \ding{109} & \ding{109}  &  \ding{109} & \ding{109}   \\ 
\hline 
Pre-activated  & \ding{109}  & \ding{52} & \ding{52} & \ding{109}  & \ding{109}  \\  
\hline 
Activated  & \ding{52} & \ding{52} & \ding{52} & \ding{52} &  \ding{109}   \\  
\hline 
Effector  & \ding{52} & \ding{52} & \ding{52} & \ding{52} & \ding{52}  \\  
\hline 
Memory  & \ding{109} & \ding{52} & \ding{52} & \ding{52} & \ding{52}  \\ 
\hline 
\end{tabular} 
\captionof{table}{Main rules applying to APCs and CD8 T-cells in the model. \ding{52}: able, \ding{109}: unable}
\label{Proprietes}
\end{center} \end{table}
 }

\subsubsection{Differentiation}\label{SectionScheme}
\begin{figure}[t !]
\centering
\includegraphics[width=14 cm]{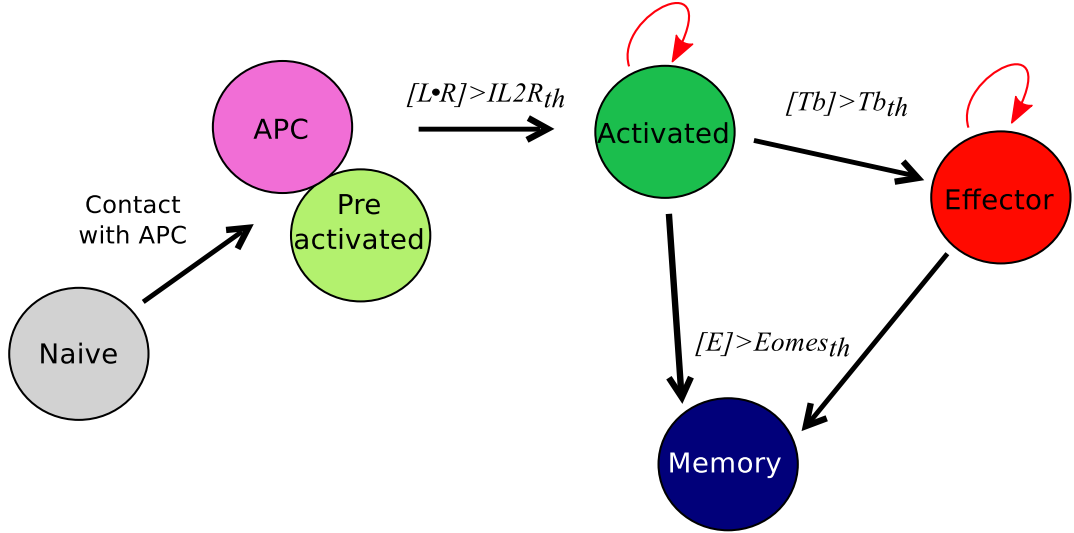}
\caption{CD8 T-cell differentiation scheme. Red arrows: proliferation; black arrows: differentiation; $th=$threshold.}
\label{Differenciation}
\end{figure}

We designed a set of rules based on the linear, irreversible differentiation scheme from \cite{Prokopiou,Gao2016}, allowing the description of a full CD8 T cell response, from the activation of naive cells up to the generation of memory cells. The differentiation pathway is illustrated in Figure \ref{Differenciation}. 

A naive CD8 T-cell binding an APC becomes pre-activated and maintains the contact with the APC thanks to good adhesion properties (cf. Section \ref{popscale} and Table \ref{paramPotts} from Appendix Section \ref{AppendixA}). If the concentration $[L\bullet R]$ of activated IL2 receptors in a pre-activated CD8 T-cell reaches a given threshold $IL2R_{th}$, the pre-activated CD8 T-cell becomes activated, leaves the APC, and starts to proliferate. When an activated CD8 T-cell divides, it gives birth to two CD8 T-cells whose states are determined by their respective concentrations of protein Tbet by comparison with a given threshold $Tbet_{th}$: activated if $[Tb]<Tbet_{th}$, effector otherwise.  Finally, if the concentration of protein Eomes is greater than the threshold $Eomes_{th}$, a dividing activated or effector CD8 T-cell will differentiate into memory cell and stop proliferating. 

\subsubsection{Cell cycle length}\label{SectionCycle}

Division is considered only for activated and effector CD8 T-cells. The cell cycle length (hours) of a cell preparing its $k$-$th$ division ($k\geq 0$) is chosen, at cell birth, from uniform law $\mathcal{U}_{[c_k-4,c_k+4]}$ where $c_k=6+28 k^2/(k^2+100)$ such that the mean duration of the cycle length increases with the number of divisions and can range from 2 to 32 hours \cite{Kinjyo2015,Yoon2010}.  At the outcome of a division, activated and effector CD8 T-cells immediately enter a new cycle.

\subsubsection{Protein distribution between daughter cells}
\label{SubsectionDistrib}

When a CD8 T-cell divides, the molecular content of the mother cell is randomly divided between the two daughter cells. To account for protein distribution between daughter cells at each division and for each protein, let us introduce the parameter $m$, defined as the \textit{degree of unevenness}. We say that divisions are $m\%$ uneven if at division one daughter cell inherits up to $(50+m/2)\%$ of the mother cell's content, while the second daughter cell receives the rest, that is at least $(50-m/2)\%$ of the mother cell's content. Then, the molecular content of each daughter cell evolves according to System (\ref{EqR})-(\ref{EqE}) until the next division. 

For the sake of clarity, we emphasise that the degree of unevenness $m$ is not the percentage of proteins received by daughters cells at each division but indicates to what extent stochastic molecular partitioning can be uneven. Based on estimation from \cite{Luzyanina2013}, we consider that divisions are $10\%$ uneven, so that the most uneven partitioning in this case would split $45\%$ and $55\%$ of the mother cell's proteins in the two daughter cells respectively. 

The exact value of each daughter cell molecular content at birth is randomly chosen according to a probabilistic law, as detailed hereafter. Each protein concentration $[i]$ of the six proteins in System (\ref{EqR})-(\ref{EqE}) is unevenly distributed among daughter cells: one cell inherits $k_i[i]$ and the other $(2-k_i)[i]$. Coefficients $k_i,\ i=1,\dots,6$, are different for each protein, each cell, and each division, and are chosen from the probability law  $\mathcal{U}_{[1-m/100,1]}$. Unless otherwise indicated, we consider $10\%$ uneven divisions \cite{Luzyanina2013}, \textit{i.e.}  $k_i\in [0.9,1]$ for $i=1,\dots,6$. One may note that $k_i\in[0,1]$ so the quantity of molecular material is preserved at each division, given that the volume of each daughter cell is half the volume of the mother.  Different degrees of unevenness will be considered in Section \ref{SectionResultAsym}. 

One special case of division is the asymmetric division, and its associated unequal repartition of Tbet between daughter cells. To account for  polarisation of naive cells by antigenic signalling and the consecutive asymmetric divisions, the first division of a CD8 T-cell following its activation by an APC is characterised by a very specific uneven distribution of protein Tbet only between the two daughter cells: a coefficient $K$ is randomly chosen from the uniform law $\mathcal{U}_{[0.5,1]}$,  one of the daughter cells is 	arbitrarily designated as the proximal daughter and receives a concentration $(2-K)[Tb]$ for protein Tbet while the other one is designated as the distal one and receives a concentration $K[Tb]$ where $[Tb]$ is the Tbet concentration in the mother cell, so that Tbet accumulates in proximal cells \cite{Chang2007,Chang2011}.  Other proteins concentrations are partitioned according to the previously mentioned rule, see paragraphs above. 

\subsubsection{Apoptosis}

CD8 T-cell apoptosis occurs as soon as Caspases concentration  $[Cas]$ reaches the threshold $Caspases_{th}$. APCs are present from the beginning of the simulation and their lifetime is randomly chosen from the uniform law $\mathcal{U}_{[48,96]}$ (hours). APCs' only role is to activate naive CD8 T-cells, so we do not model any molecular activity within APCs. Dead cells are removed from the domain.

\subsection{Spatial modeling and cellular interactions}
\label{popscale}
At the cell population scale, we use a cellular Potts model (CPM), also known as Glazier-Graner-Hegeweg model \cite{GGH}, to describe a population of CD8 T-cells and APCs evolving in a two-dimensional domain. Basically, a CPM is a time-discrete algorithm where cells, or agents, are defined as sets of nodes and move on a lattice, one node at a time, according to probabilistic rules based on the minimisation of the energy of the system, known as the Hamiltonian.
 
In our model, based on that from \cite{Prokopiou,Gao2016}, the domain is a square lattice of $S=150\times 150$ nodes with periodic boundary conditions. Each node $\vec{x}$ bears an index $\sigma(\vec{x})$. A set of nodes bearing the same index $\sigma$ defines a cell, also denoted by $\sigma$. Finally, each cell $\sigma$ has a type $\tau( \sigma)$ defining its properties. In our case, the different types are: extracellular medium, APC, naive, pre-activated, activated, effector and memory CD8 T-cell. Note that, technically, the extracellular medium is considered as a cell, denoted by $\sigma_e$.

Cell (including extracellular medium) size variation and displacement result from the succession of copies of index from nodes to neighbour nodes, based on the minimisation of the Hamiltonian $\Omega$ (see equation (\ref{Hamiltonien})), thanks to a simulated annealing algorithm. More precisely, at each iteration, known as Monte Carlo Step (MCS), of the CPM, the following algorithm is executed $N=3\times S$ times:\begin{description}
\item[Step 1 ] Randomly choose a source node $x_s$ and, among its first order neighbours, a target node $x_g$.
\item[Step 2 ] Compute the Hamiltonian $\Omega$, and the putative Hamiltonian $\Omega'$ that would be obtained if node $x_s$ would copy its index on node $x_g$, \textit{i.e.} if cell $\sigma(x_s)$ incorporates the node $x_g$.
\item[Step 3 ] Compute $\Delta \Omega= \Omega-\Omega'+\Delta_{\text{motility}}$ (see equation (\ref{DeltaMobilite}) below) to evaluate the energy cost of such a copy.  If $\Delta \Omega>0$, $x_s$ copies its index $\sigma(x_s)$ on $x_g$, \textit{i.e.} $x_g$ is integrated by cell $\sigma(x_s)$. Else, the copy is accepted with probability $\exp(-\Delta \Omega/T)$, known as Boltzman probability, where parameter $T$ characterises the propensity of the system to evolve.
\end{description}
Note that it is conventional to consider $N=S$ pixel copy attempts per MCS. However, in that case the maximum speed cells can reach is limited to approximatively 0.1 pixel per MCS \cite{Swat2012}, which eventually defines a finer time resolution than expected for the integration of differential equations. We emphasise that this limitation can be removed by increasing this number (here $N=3\times S$).

The Hamiltonian $\Omega$ is computed using the following formula:
 \begin{align}
\Omega =\ & \underset{\text{perimeter}}{\underbrace{ \lambda_{pm}\underset{\sigma \neq \sigma_e}{\Sigma}(p_{\sigma}-P_{\tau(\sigma)})^2} } +
\underset{\text{area}}{\underbrace{ \lambda_{area}\underset{\sigma \neq \sigma_e}{\Sigma}(a_{\sigma}-A_{\tau(\sigma)})^2}}\nonumber \\   &+
\underset{\text{contact}}{\underbrace{ \underset{\text{neighbours }(\vec{x},\vec{x^*})}{\Sigma}J_{\tau (\sigma (\vec{x})),\tau(\sigma(\vec{x^*}))}(1-\delta_{\sigma(\vec{x}),\sigma(\vec{x^*})}) }},   
\label{Hamiltonien}
\end{align} 
where $J_{\tau_1,\tau_2}$ accounts for the contact energy between two cells of types $\tau_1$ and $\tau_2$. Thanks to the term  $1-\delta_{\sigma(\vec{x}),\sigma(\vec{x^*})}$, two neighbour nodes belonging to the same cell do not generate contact energy.
$p_\sigma$ and $a_\sigma$ are the actual perimeter and area of cell $\sigma$, respectively, whereas $P_{\tau(\sigma)}$ and $A_{\tau(\sigma)}$ are the target perimeter and area, respectively, for a cell of type $\tau(\sigma)$ ; perimeter and area constraints then penalize the configurations where the effective perimeter and area are distant from the target ones. Parameters $\lambda_{area}$ and $\lambda_{pm}$ define the weights of those two constraints. The perimeter constraint has been added to the definition used in \cite{Prokopiou,Gao2016} in order to avoid potential cell fragmentation.

The energy $\Delta_{\text{motility}}$ is defined by
\begin{equation}
\Delta_{\text{motility}}=v(\sigma(x_s))  \left(cos(\theta(\sigma(x_s),t)),sin(\theta(\sigma(x_s),t))\right)\cdot (x_g-x_s),
\label{DeltaMobilite}
\end{equation}
where $v(\sigma(x_s))$ is the weight associated to the motility energy for the cell $\sigma(x_s)$ and $\theta(\sigma(x_s),t)$ is the privileged angle of direction for the cell $\sigma(x_s)$ at time $t$, randomly updated along the simulation. The operator ``$\cdot$" stands for the dot product. Thus, $\Delta_{\text{motility}}$ is all the more high (and then the copy is all the more probably accepted) that the copy direction $(x_g-x_s)$ aligns with $(cos(\theta(t)),sin(\theta(t)))$.

\subsection{Numerical resolution}\label{SectionResolution}
The initial cell population is composed of 30 naive CD8 T-cells and 3 APCs. A simulation requires 30,000 iterations (MCS) corresponding to 20 days and 20 hours in the real time, that is, 1 MCS represents 1 minute. When a simulation starts, APCs are already present in the LN, ready to activate naive CD8 T-cells. We consider the initial time to be day 4 post-infection (D4 p.i.) since our \textit{in vivo} data set starts D4 p.i.. 

We assume that a node of the lattice corresponds to $4\times 4 \mu m^2$ for biological interpretation. The target cell area is chosen to be 9 nodes ($144\mu$m$^2$) for CD8 T-cells and 140 nodes ($2240\mu$m$^2$) for APCs. The target perimeter for CD8 T-cells is $48 \mu m$ in order to favour compact shapes ; there is no constraint on APC perimeter.  The simulations have been performed using CC-IN2P3 servers on Compucell3D software \cite{Swat2012} with, unless otherwise stated, the parameter values from Tables \ref{paramPotts}, \ref{paramIntra}, \ref{paramIL2} and \ref{paramThresholds}  from Appendix Section \ref{AppendixA}.%

In Section \ref{SectionMemory}, we study the ability of our model to simulate a secondary response, also called memory response. Our model has first been calibrated in order to reproduce an \textit{in vivo} primary response  against \textit{Listeria monocytogenes} (\textit{Lm}) infection from \cite{Badovinac2003} (see Figure \ref{ResultsBadovinac}). Then, the same parameter values have been used to simulated both a primary and secondary responses. However, secondary response simulations are performed with initially 3 APCs and 30 memory CD8 T-cells (instead of 30 naive CD8 T-cells for the primary response) that are able to bind an APC to become pre-activated, then the differentiation scheme presented in Section \ref{SectionScheme} applies. The molecular profile of the initial memory cells is set as the asymptotic molecular profile developed by memory cells at the end of a primary response, as discussed in Section \ref{SectionSLECMPEC}.  

\subsection{Model calibration}
Parameters of equations (\ref{EqR})-(\ref{DeltaMobilite}) have been calibrated on \textit{in vivo} data using parameter values from  \cite{Prokopiou,Gao2016}. Since handling big cell populations with an agent-based model implies expensive computation time, we focused on fitting the proportion, rather than the number, of CD8 T-cells in each state of differentiation among the whole cell population.  In order to compare \textit{in silico} and \textit{in vivo} data at both cellular and molecular scales we minimised the metric 
$D=D_{cell}+D_{prot}$ where
 \begin{equation}
D_{cell}=\frac{1}{(\#S)(\#V)}\underset{\text{simulation}\ \mathcal{S}}{\sum}\cdot \underset{\text{mouse}\ \mathcal{V}}{\sum}\cdot \underset{\text{cellular type}\ \mathcal{C}}{\sum}\cdot \underset{\text{time step} \ t}{\sum}|\mathcal{S}(\mathcal{C},t)-\mathcal{V}(\mathcal{C},t) |
\label{Dcell} 
\end{equation} and\begin{equation}
D_{prot}=\frac{1}{(\#S)(\#V)}\underset{\text{simulation}\ \mathcal{S}}{\sum}\cdot \underset{\text{mouse}\ \mathcal{V}}{\sum}\cdot \underset{\text{protein}\ \mathcal{P}}{\sum}\cdot \underset{\text{time step} \ t}{\sum}|\mathcal{S}(\mathcal{P},t)-\mathcal{V}(\mathcal{P},t) |,
\label{Dprot}\end{equation} 
with 
 $\#S$ the number of simulations performed with a given set of parameters and $\#V$ the number of mice from which \textit{in vivo} data have been collected. $\mathcal{S}(\mathcal{C},t)$ (resp. $\mathcal{V}(\mathcal{C},t)$) is the ratio between the number of cells of type $\mathcal{C}$ and the size of the CD8 T-cell population at time $t$ in the simulation $\mathcal{S}$ (resp. the mouse $\mathcal{V}$). $\mathcal{S}(\mathcal{P},t)$ (resp. $\mathcal{V}(\mathcal{P},t)$) is the ratio between the mean concentration  of protein (resp. expression of mRNA) $\mathcal{P}$ among the CD8 T-cell population at time $t$ in the simulation $\mathcal{S}$ (resp. the mouse $\mathcal{V}$) and the maximal concentration (resp. expression) observed among all the time steps.

Since pre-activated and activated cellular types are not identified in \textit{in vivo} data, we gathered pre-activated with naive T-cells and activated with effector T-cells. Then cellular types $\mathcal{C}$ in equation (\ref{Dcell}) are:  naive/pre-activated, activated/effector and memory.  In equation (\ref{Dprot}), quantities $\mathcal{P}$ are the ones for which we have relevant \textit{in vivo} mRNA expression data at our disposal: IL2 receptors, Tbet and Eomes. 

Note that we did not perform a parameter estimation procedure, but a calibration of our model based on experimental data. Evaluation of accuracy and sensitivity of parameter values have been investigated in previous studies \cite{Gao2016,Prokopiou}. Since we modified the model to account for differentiation in memory cells, a sensitivity analysis of our model to parameter $Eomes_{th}$ is presented in Appendix Section \ref{AppendixB}.

\section{Results}
\label{SectionResults}

\subsection{modeling the CD8 T-cell immune response at both cellular and molecular scales} \label{SectionResultsData}
\begin{figure}[tp!]
\centering
\includegraphics[width=10 cm]{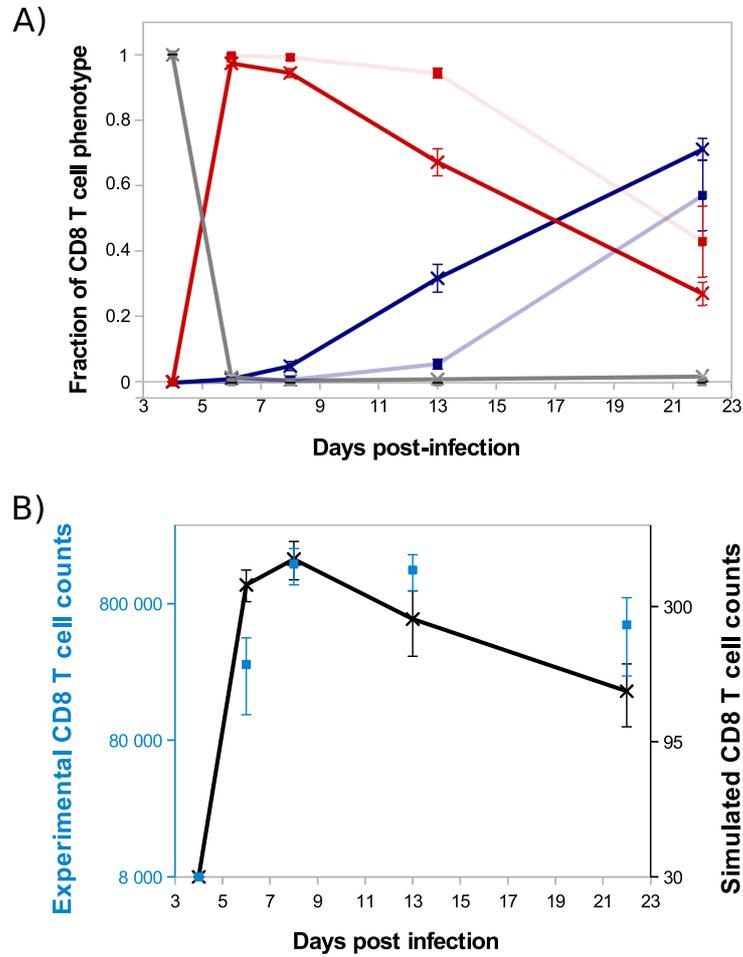}
   \caption{CD8 T-cell population dynamics. (A) Fraction of each cell type among the CD8 T-cell population. Grey: naive+pre-activated cells; red: activated+effector cells; blue: memory cells; full lines with crosses: \textit{in silico} (mean +/- standard deviation over 10 simulations); transparent lines with squares: \textit{in vivo} (mean +/- standard deviation over data from 5 mice). Error bars are most of time very small and then not visible.
(B) Size of the CD8 T-cell population \textit{in silico} (black crosses, right y-axis, mean +/- standard deviation over 10 simulations) and \textit{in vivo} (blue squares, left y-axis, mean +/- standard deviation over data from 5 mice). } 
   \label{ResultsPop}
\end{figure}

We first briefly illustrate our model's ability to reproduce \textit{in vivo} dynamics at both cellular and molecular scales. The evolution of the composition of a CD8 T-cell population from D4 to D22 p.i. is presented on Figure  \ref{ResultsPop}.A. In both  \textit{in vivo} and \textit{in silico} data, naive CD8 T-cells are negligible after D6 p.i.. At the peak of the response, occurring D8 p.i. both  \textit{in vivo} and \textit{in silico}, more than $94\%$ of the CD8 T-cells are in the activated or effector state, while the memory population emerges during the subsequent contraction phase. As a result of effector cell death and differentiation, memory cells represent the major part of the population on D22 p.i..  Figure \ref{ResultsPop}.B shows the size, in number of cells, of the CD8 T-cell population.  The qualitative \textit{in vivo} dynamics is quite well reproduced: antigen presentation to naive CD8 T-cells triggers clonal expansion, population size reaches a peak D8 p.i. followed by a contraction phase where most cells ($64\%$ and $67\%$ \textit{in vivo} and \textit{in silico} respectively) die.  

\begin{figure}[tp!]
\centering
 \includegraphics[width=8 cm]{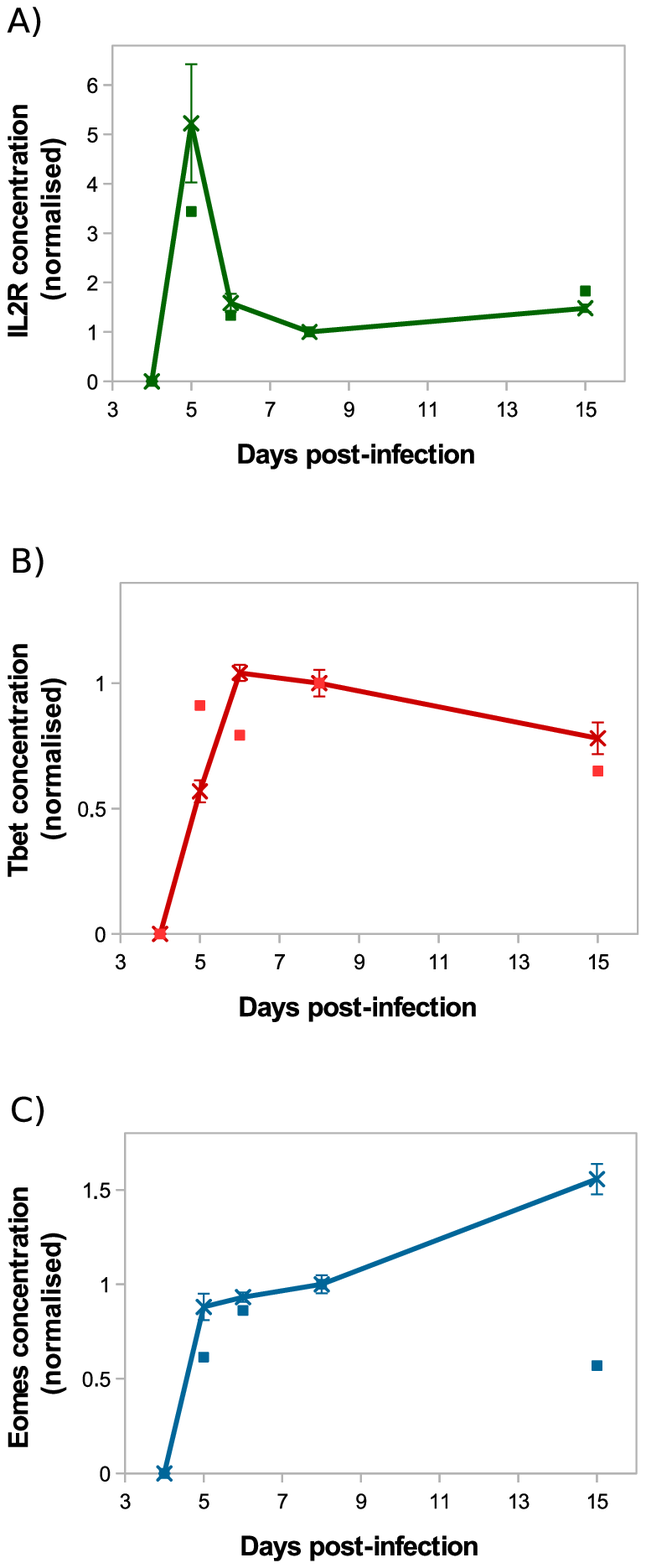}
   \caption{Molecular dynamics. Mean concentration of (A) both activated and inactivated IL2 receptors, (B) Tbet  and (C) Eomes among the CD8 T-cell population normalised by the concentration value  D8 p.i..  Lines with crosses: \textit{in silico} (mean +/- standard deviation over 10 simulations); squares: \textit{in vivo} mRNA data from ImmGen.}
   \label{ResultsMolec}
\end{figure}
On Figure \ref{ResultsMolec}, \textit{in silico} predictions are compared to the mean IL2 receptors, Tbet and Eomes mRNA expression levels of CD8 T cells activated in vivo. The kinetics of IL2R and Tbet are well reproduced. Indeed, as a result of TCR engagement, IL2R concentration sharply increases and reaches a peak D5 p.i., allowing cells to capture IL2 and get activated. Then IL2R concentration decreases until D8 p.i. and slowly increases from D8 to D15 p.i.. Tbet concentration increases from D4 to D6 p.i. and remains stable until D8 p.i., then decreases until D15 p.i.. Mean Tbet concentration consistently correlates with the size of effector CD8 T-cell population (Figures \ref{ResultsPop}.A and \ref{ResultsMolec}.B) and is in agreement with its role in the control of cytotoxicity and cell apoptosis. Regarding Eomes concentration, the \textit{in vivo} increase between D4 and D8 p.i. is well reproduced by our model, however the increase observed between D8 and D15 p.i. does not match the \textit{in vivo} data. As cells evolve towards a memory phenotype, \textit{in silico} Eomes concentration increases and up-regulates the expression of IL2R (Figure \ref{ModeleBio}) to exacerbate the sensitivity of memory cells to IL2. It should be noted that various works support that Eomes expression increases in effector cells progressing toward a memory phenotype \cite{Lazarevic2013,Joshi2011,McLane2013}, contrary to what is observed in the mRNA dataset from Immgen.

\subsection{Cellular dynamics arise from cellular heterogeneity}
\label{SectionSLECMPEC}
In our model, each cell develops its own molecular profile, resulting in a heterogeneous cell population. Consequently, studying the mean concentration of a given protein among the population, as shown on Figure \ref{ResultsMolec} for example, is not sufficient to understand the molecular dynamics among the CD8 T-cell population.

\begin{figure}[tp!]
\centering
 \includegraphics[width=14cm]{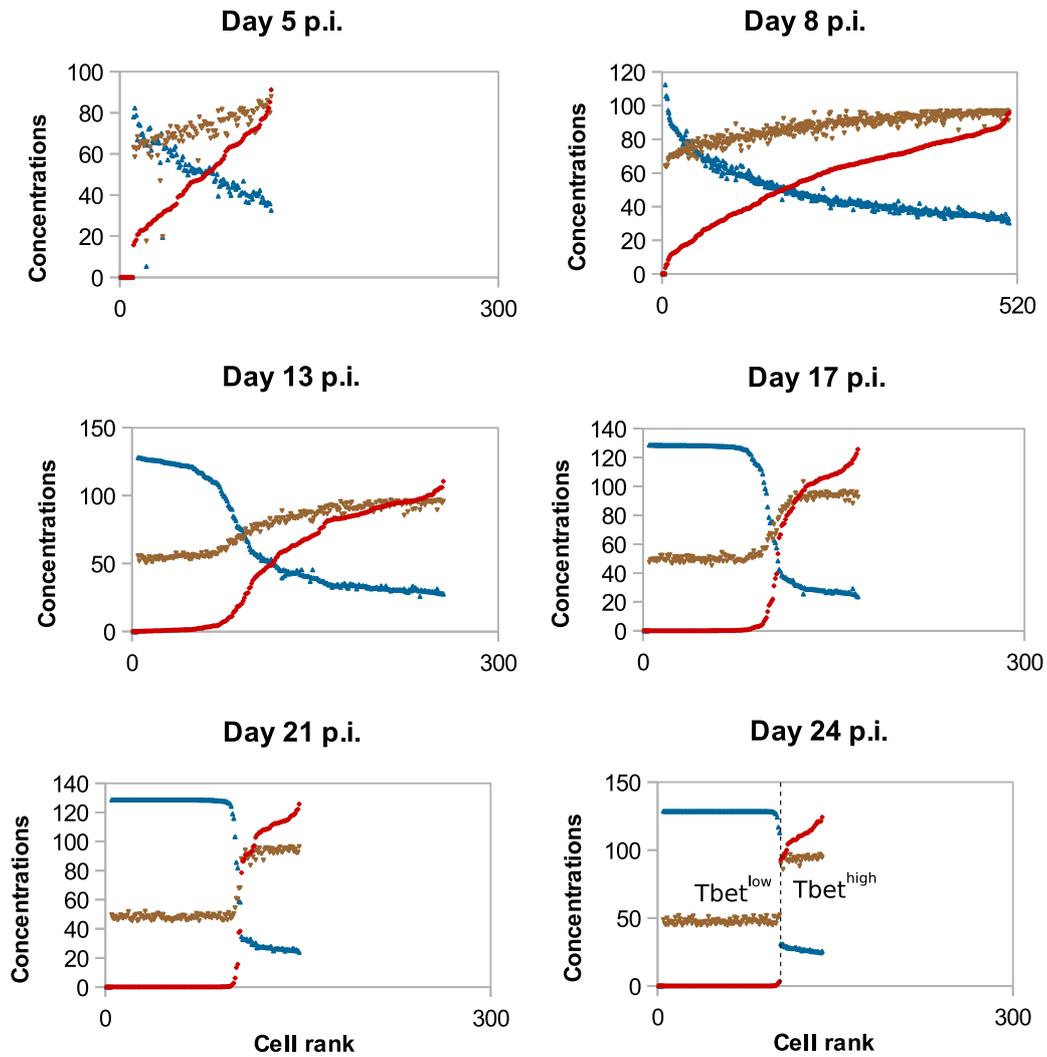}\\
   \caption{Concentrations (mol/L) of Tbet (red), Eomes (blue) and Caspases (brown) in all cells, sorted (left to right) according to their Tbet concentration. \textit{In silico} CD8 T-cell population on D5, D8, D13, D17, D21 and D24 post-infection are represented. To make it easier to read,  Eomes and Caspases concentrations have been multiplied by factor 5.}
   \label{ResultsConcentration}
\end{figure}

To study the molecular-content heterogeneity and its role in cellular dynamics, we show in Figure \ref{ResultsConcentration} the \textit{in silico} concentrations of Tbet, Eomes and Caspases in each CD8 T-cell of the population at different times of the response. Cells were ranked according to their Tbet content.  D5 to D8 p.i., corresponding to the clonal expansion phase (see Figure \ref{ResultsPop}), concentrations are heterogeneous but uniformly distributed around the mean value. Most of that heterogeneity comes from the conditions of activation and from molecular partitioning at cell division. Yet from D8 to D24 p.i., corresponding to the contraction phase, two sub-populations of cells clearly emerge: one with high concentration of Tbet (centred around $[Tb]_s\approx 118$ mol/L) and one with low concentration of Tbet ($\approx 0$ mol/L). The unstable steady state of (\ref{EqTb}), defined in Proposition \ref{PropTbet}  and separating the stable equilibria $0$ and $[Tb]_s$, is given by $[Tb]_u\approx 21$ mol/L. Moreover, cells expressing high levels of Tbet express high levels of Caspases and low levels of Eomes, a molecular profile associated with cell death and poor memory potential. On the contrary, cells expressing low levels of Tbet have good survival and memory differentiation properties since they express low levels of Caspases and high levels of Eomes. Progressively, cells with high concentrations of Tbet die (when their concentrations of Caspases reach the threshold $Caspases_{th}\approx 19$ mol/L) and cells with low concentrations of Tbet differentiate into memory cells and stop proliferating (when the concentration of Eomes reaches $Eomes_{th}=16$ mol/L). On D24 p.i. there is no cell with intermediary profile, most of the cells have differentiated into memory cells while a few effector cells with high Tbet concentrations still survive. One can observe that the molecular profiles of memory cells  converge to the same state where $[Tb]=0$ mol/L, $[E]\approx 26$ mol/L and $[Cas]\approx 9$ mol/L. 

The coexistence of two sub-populations  characterised by their concentrations of Tbet explains the population dynamics observed on Figure \ref{ResultsPop}. That is, the contraction of the cytotoxic effector cell population simultaneously with the emergence of a memory cell population with survival properties. 

As discussed in the introduction, responding CD8 T-cells can be distinguished between short-lived (SLEC) and memory precursor (MPEC) effector cells based on the expression of two proteins: KLRG1 and CD127 \cite{Blueprint,Joshi2011}. In this section, we investigated how, in our model, the heterogeneity of Tbet concentrations among a CD8 T-cell population explains the emergence of two sub-populations of CD8 T-cells. The first one, expressing high concentrations of Tbet, could be comparable to SLEC that exhibit properties such as apoptosis and cytotoxicity, a process regulated by Tbet. The second one (memory potential, survival)  would be similar to the MPEC population.   This is consistent with the litterature, since Tbet is known to favour the development of SLEC, to the detriment of MPEC \cite{Joshi2007,Huang2015,Lazarevic2013}.

\subsection{Moderate uneven molecular partitioning favours efficient generation of memory cells}\label{SectionResultAsym}
\begin{figure}[tp!]
\centering
 \includegraphics[width=14cm]{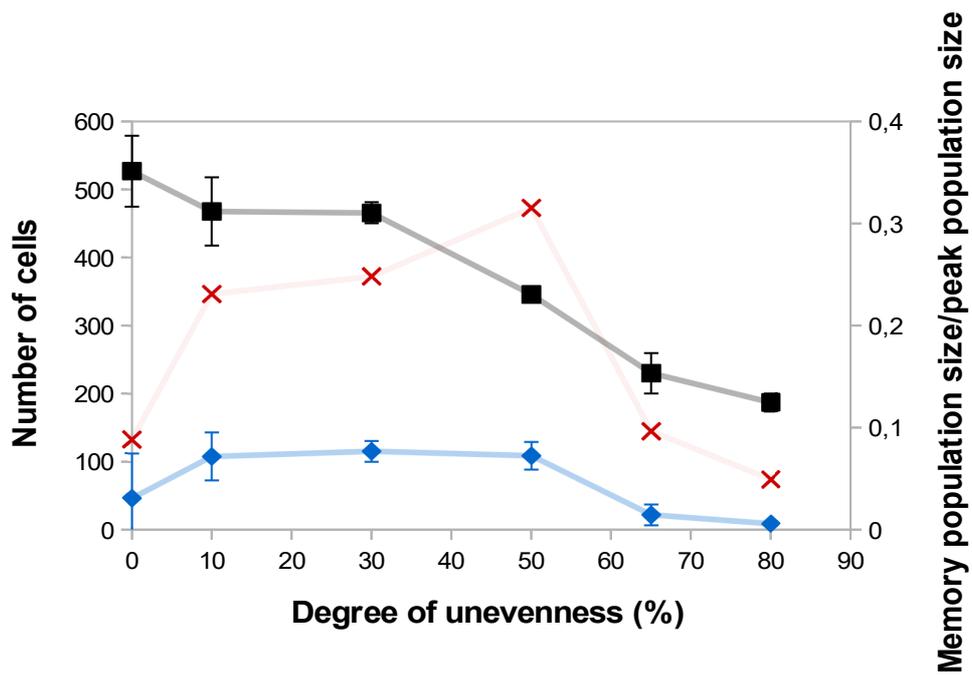}\\
   \caption{Size of the CD8 T-cell population at the peak of the response (black squares, left axis) and size of the memory CD8 T-cell population at the end of the response D25 p.i. (blue diamonds, left axis) as functions of the degree  unevenness of molecular partitioning (mean +/- standard deviation over 5 simulations).   
   Red crosses (right axis) show memory cell generation efficiency, measured as the ratio between the size of the memory CD8 T-cell population D25 p.i. and the size of the CD8 T-cell population at the peak of the response (mean over 5 simulations).}
   \label{ResultsAsym}
\end{figure}

A major source of heterogeneity in our model is the uneven molecular partitioning at cell division determined by the degree of unevenness $m$ (see Section \ref{SubsectionDistrib}). We compare on Figure \ref{ResultsAsym} the sizes of the CD8 T-cell population at the peak of the response  as well as the sizes of the memory population on D25 p.i. for different degrees of unevenness, that is the extent of unevenness of the stochastic molecular partitioning. We do not however modify the degree of unevenness of the asymmetric first division, consecutive to the polarisation of the cell due to APC binding  \cite{Chang2007,Chang2011}, see Section \ref{SubsectionDistrib}.

First, Figure \ref{ResultsAsym} shows that the size of the CD8 T-cell population at the peak of the response decreases as the degree of unevenness increases.  Indeed, the more uneven the molecular partitioning, the sooner CD8 T-cells expressing high levels of Caspases or Eomes appear and then the sooner cells die by apoptosis or differentiate in non-proliferating memory cells.

 Second, the relation between the degree of unevenness and the size of the memory population generated at the end of the response is not monotonous: the biggest memory populations are observed when considering a moderate unevenness ($10\%$-$50\%$). 
 
In Section \ref{SectionSLECMPEC}, the role of Tbet concentration in determining the fate  (death or memory differentiation) of an effector CD8 T-cell has been discussed. Additionally, we showed in \cite{Girel2018} that the progression of a cell lineage towards death or memory differentiation can be slowed down or reversed by molecular partitioning depending on cell cycle length, initial Tbet concentration and the degree of unevenness.  This stressed, on a simplified model, the influence of the degree of unevenness on cell fate choice regulation. 

On the opposite, when molecular partitioning is symmetrical ($m=0$) and no further T-cell-APC interactions are assumed, there is no more source of stochasticity and consequently all the CD8 T-cells of the same lineage express the same concentration of Tbet. As a consequence of Proposition \ref{PropTbet}, this concentration irreversibly converges either to $[Tb]_s$ (high Tbet concentration) or to $0$ mol/L (low Tbet concentration). This irreversibly leads to apoptosis (high Tbet concentration) or memory differentiation (low Tbet concentration) of the whole cell lineage.  

Thus, our result clearly stresses that uneven partitioning allows the maintenance of a CD8 T-cell compartment with undetermined fate for some time, through cell fate reversibility. As long as it is maintained, this compartment is able to produce both effector cells destined to die and memory cells. 

We also showed in \cite{Girel2018} that the higher the degree of unevenness, the more reversible the cellular fate. Surprisingly, strong unevenness ($65\%-80\%$) results in smaller memory cell populations (Figure \ref{ResultsAsym}). In fact, strong unevenness favours the fast emergence of daughter cells with very high or  low concentrations of Tbet such that those cell lineages are likely to die or to generate memory cells. In particular, effector cells with high memory potential poorly expand before they differentiate hence this leads to the generation of fewer memory cells.

To discuss the efficiency of memory cell generation, we compare on Figure \ref{ResultsAsym} the number of memory cells generated at the end of the response to the number of cells at the peak of the response, viewed as an indicator of the energetic cost of the response for the organism (red crosses). Figure \ref{ResultsAsym} suggests that the degree of unevenness in molecular partitioning impacts memory generation, with the better ratio (more than $30\%$) obtained when considering 50$\%$ uneven molecular partitioning.

\subsection{Memory response}\label{SectionMemory}
One of the characteristics of memory cells is their capacity to mount more rapid effector response than naive cells and to generate an increased fraction of memory cells \cite{Wirth2010}. To test whether the memory cells generated by our model exhibit some of these features we compared the \textit{in silico} primary response with a secondary response of \textit{in silico} generated memory cells.

\begin{figure}[tp!]
\centering
 \includegraphics[width=17cm]{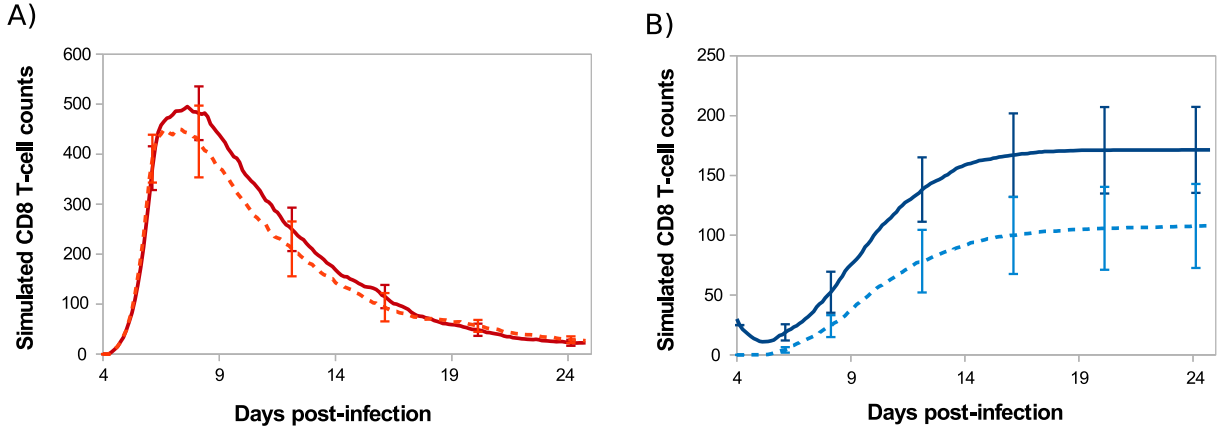}
   \caption{Number of (A) activated/effector and (B) memory CD8 T-cells during  \textit{in silico} primary (dashed line) and secondary (full line) responses. Mean +/- standard deviation over 10 simulations. }
   \label{ResultsMemory}
\end{figure}

Figure \ref{ResultsMemory} shows the \textit{in silico} memory response (or secondary response), obtained with an initial population of 30 memory T-cells, as described in Section \ref{SectionResolution}. This secondary response is compared to the primary immune response  starting with 30 naive CD8 T-cells (Section \ref{SectionResultsData}). The \textit{in silico} secondary response is characterised by a bigger CD8 T-cell population, at any time of the response. From the primary to the secondary response, there is a small increase in the size of the sub-population of activated and effector cells but the major change is in the size of the memory population. Indeed, the number of memory CD8 T-cells increases much faster during the secondary response such that D29 p.i. the memory population is two times bigger than during the primary response.
 This can be explained by the fact that memory cells are activated faster than naive cells, thanks to their molecular profile. Indeed, memory cells express higher concentrations of IL2 receptors than naive cells,  since it is  sustained by the expression of Eomes. Consequently, the threshold $IL2R_{th}$ (see Section \ref{SectionScheme}) is reached sooner  when starting with memory cells than with naive cells. As a result, the concentration of Tbet, up-regulated during APC binding, is lower after the activation of a memory cell than after the activation of a naive cell, and low Tbet level is associated to memory precursor fate and low cytotoxicity.

\begin{figure}[tp!]
\centering
 \includegraphics[width=13cm]{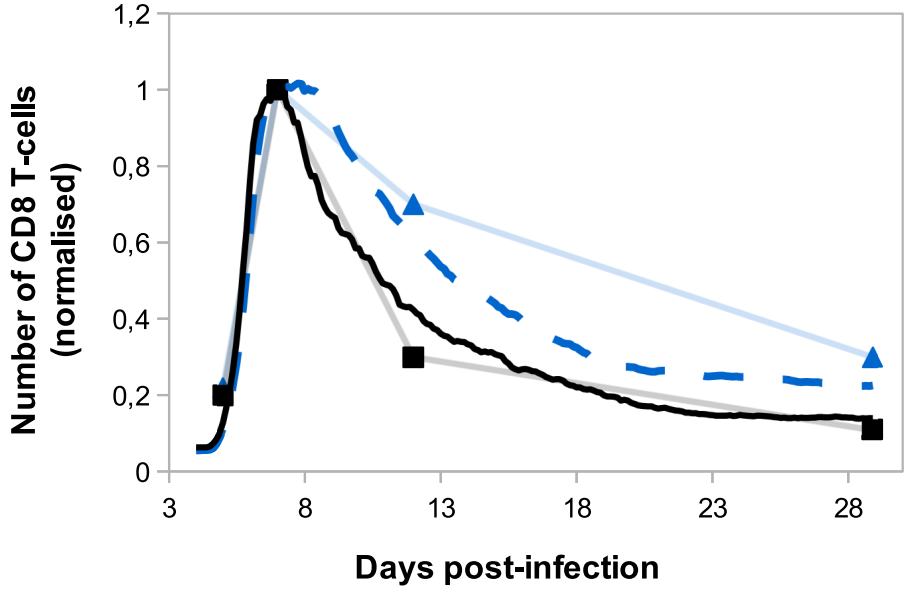}
   \caption{Number of  CD8 T-cells, normalised by CD8 T-cell population size D7 p.i.,  during  \textit{in silico} primary (black full line) and secondary (blue dashed line) responses (mean   over 10 simulations) compared with \textit{in vivo} primary (black squares) and secondary (blue triangles) responses against \textit{Listeria monocytogenes} from \cite{Badovinac2003}. }
   \label{ResultsBadovinac}
\end{figure}

On Figure \ref{ResultsBadovinac}, we compare \textit{in silico} primary and secondary responses from our model with \textit{in vivo} primary and secondary responses against \textit{Lm} infection from \cite{Badovinac2003}.  Since our model has been calibrated to fit the primary response data, we do not aspire to reproduce the quantitative dynamics of the secondary response, but rather to study its qualitative properties. Namely, the secondary response is characterised by a slower and weaker contraction phase, from the peak of the response D7 p.i. to the last time point D29 p.i.. This weaker contraction could be explained by an early production of memory cells that leads to a large population of memory cells, as it is the case in our model (Figure \ref{ResultsMemory}). 
\section{Discussion}
Activation of naive CD8 T-cells triggers a primary immune response, characterised by a well orchestrated program of cell proliferation, differentiation, death and migration. It is now well known that the responding CD8 T-cell population is heterogeneous and that a single naive T-cell can generate differently fated cells \cite{Gerlach2010}. However, evaluating how cellular and molecular events contribute to that heterogeneity and identifying its consequences  on the outcomes of the immune response  remain fundamental questions.

With this in mind, we expanded a hybrid  multi-scale model of the CD8 T-cell immune response, where cell behaviour is determined by intracellular molecular dynamics.  Model parameters have been calibrated using \textit{in vivo} data at both cellular and molecular scales. Because of expensive running time, we were led to simulate small cell populations so that we focused on semi-quantitative fitting criteria. After calibration, our model succeeded  in reproducing the temporal dynamics of the response regarding the size of the CD8 T-cell population and the proportion of cells in each differentiation stage.  Apart from a discordance between \textit{in silico}  and \textit{in vivo} mean concentration of Eomes on day 15 p.i., our model captured the dynamics of the mean concentration of IL2 receptors, Tbet and Eomes, which play key roles in the differentiation processes.
 
In addition to reproduce primary responses, our model easily produces secondary responses. Memory cells generated during the \textit{in silico} primary response succeeded in mounting a stronger secondary response upon antigenic stimulation (Figure \ref{ResultsMemory} and Figure \ref{ResultsBadovinac}). It should be noted that the differences between outcomes of the primary and secondary \textit{in silico} immune responses only depend, in this work, on the difference between the molecular profile of memory and naive CD8 T-cells and do not take into account a lot of characteristics of the secondary response described in the literature such as: biggest initial CD8 T-cell population \cite{Blattman2002}, shorter cell cycle \cite{Veiga2000} or sensitivity to inflammatory cytokines, such as IL12 \cite{Joshi2011}.

 We discussed how a deterministic description of molecular concentration dynamics combined with stochastic events, such as uneven partitioning of molecular content at division, can regulate the emergence and the maintenance of two sub-populations of CD8 T-cells. Those sub-populations, characterised by their molecular profiles, coexist but express different properties and antagonistic fates, comparable to those of SLEC and MPEC described in the literature \cite{Joshi2007}.  From that observation, we showed that the dynamics observed at the cellular scale (cell differentiation, population size) could be explained by molecular-content heterogeneity among the cell population, which mostly originates from  uneven partitioning of molecular content. We did not however consider the effect of stochastic fluctuations of gene expression,  known to be an important source of heterogeneity \cite{Raser2005}. Interestingly, Huh and Paulsson \cite{Huh2010} showed that both stochastic gene expression and stochastic partitioning of molecular content are equally good to explain the heterogeneity observed at cell division and suggested that much of the heterogeneity usually attributed to the former actually results from the latter. 
 
 In our model, cell phenotypic heterogeneity, associated with molecular-content heterogeneity, first arises upon asymmetric division of polarised naive cells. This heterogeneity is thereafter continuously regulated throughout the whole response by means of uneven partitioning of molecular-content at each division. This is in agreement with the observations of Lema\^{i}tre \textit{et al.} \cite{Lemaitre2013} who state that T-cell diversification is a continuous process, spread over the whole response, including the asymmetric first division and late events occurring throughout subsequent divisions. Besides, Lema\^{i}tre \textit{et al.} \cite{Lemaitre2013} pointed out that cellular heterogeneity, that could result from variations in naive T-cell responsiveness to cytokines or TCR signalling, pre-exists prior to the first division. In this article we did not consider preexisting heterogeneity among the naive T-cell pool, that could be achieved by varying the parameter values of System (\ref{EqR})-(\ref{EqE}) associated to each naive T-cell. We can expect that it would confer to each naive T-cell a predisposition to engender a cell lineage oriented toward either apoptosis or memory differentiation. Moreover,   the initial heterogeneity among naive T-cells could be conserved through the response, then leading to a heterogeneous pool of memory cells, a feature that is not reproduced by our model \cite{Kaech2007}.

Polarisation of naive cells upon antigenic stimulation has been observed in CD4 T-cells \cite{Pham2014,Jung2014}  and B-cells \cite{Pham2014,Barnett2011,Thaunat2012}.  This polarisation results in   asymmetric division of naive cells and may induce heterogeneous cell fates \cite{Jung2014,Barnett2011,Thaunat2012}. Regarding the subsequent divisions, it can be thought that they are subject to uneven and random partitioning of molecular content since this phenomenon has been reported in many types of cells, including   yeast, bacteria and T-cells \cite{Block1990,Bocharov2013,Golding2005,Huh2010,Luzyanina2013,Sennerstam88}. However, the contribution of uneven and random partitioning of non-polarised cells in the development of heterogeneous cell fate has not been studied yet. To that end, it would be interesting to extend the approach developed in our study to the differentiation of other lymphocytes, such as B-cells or CD4 T-cells.

In our study, increasing the degree of unevenness of molecular partitioning reduces the expansion size of the whole CD8 T-cell population whereas the size of the sub-population of memory cells is maximal for intermediate degrees of unevenness. As a consequence, the ratio between the number of memory cells generated and the magnitude of the response at its peak, viewed as a measure of memory generation efficiency, is maximised when considering a $50\%$ degree of unevenness. As discussed above, molecular partitioning is not the only regulator of heterogeneity. In this regard, we can believe that our evaluation overestimates the value of this optimal degree of unevenness and rather indicates that generating a moderate heterogeneity all along the immune response leads to efficient memory generation. 

In our manuscript, when the degree of unevenness is $m=10\%$, each daughter cell inherit from 45 to 55$\%$ of the mother cell's molecular content, with uniform probability distribution. The unevenness of molecular partitioning remains difficult to measure experimentally. Based on in vitro experimental data of CFSE dye expression, Luzyanina \textit{et al.} \cite{Luzyanina2013} estimated that the two daughter cells inherit of 42,3$\%$ and 57,7$\%$ of the mother cell's molecular content, respectively. Rather than considering a uniform probability distribution and a degree of unevenness, we could  consider that the molecular partitioning is a binomial phenomenon \cite{Golding2005}, \textit{i.e.} each protein has the same probability  to be attributed to each daughter cell. Such a discrete distribution can be approximate by a continuous and truncated (to avoid negative values) normal distribution whose variance would characterise the level of unevenness.


 Note that, in works dealing with the CD8 T-cell immune response, it is usual to consider that 5 to 10$\%$ of the cells present at the peak of the response survive the contraction phase and differentiate into memory cells \cite{Blattman2002}. This is consistent with our results only for symmetric divisions or for divisions with high ($65\%$ to $80\%$) degrees of unevenness. However, this hypothesis can be challenged, as pointed out in \cite{Crauste2017}, as for the actual \textit{in vivo} data presented in Figure \ref{ResultsPop},  D22 p.i. the memory population size is $19.5\%$  of the whole population at the peak of the response, D8 p.i.. This suggests that the amplitude, and possibly the kinetics, of the cellular contraction is not only an inherent feature of the CD8 immune response but also depends on external factors such as inflammatory factors.

In many mathematical models of the CD8 T-cell immune response, as those referenced in \cite{ANTIA2003}, cell proliferation and differentiation depend on the amount of pathogen, in the manner of prey-predator models used in ecology. In our model a brief initial antigenic stimulation of  naive CD8 T-cells is sufficient to  trigger an autonomous program of proliferation and differentiation, as stated in the literature \cite{Stipdonk,Kaech2001,Wherry2004}. However, while dispensable, \textit{in vivo} inflammatory signals can affect the immune response outcome \cite{Bevan2001}. A motivating perspective  is to evaluate  the respective contributions  of both the autonomous program and extrinsic inflammatory factors to the immune response, so that the latter could be tuned by mastering the inflammatory environment. For example, extending our model by incorporating the inflammatory cytokine IL12, secreted by APCs, could  markedly affect the effector/memory cell balance since IL12 is known to respectively promote and repress Tbet and Eomes synthesis \cite{Joshi2007,last,Takemoto2006}. 


Cell cycle length depends in our model on the number of divisions the cell has undergone.  It would be instructive to introduce a molecular control of cell proliferation, since the putative existence of coexisting sub-populations with disparate cycle lengths could considerably impact the cellular dynamics. One could for instance consider the transcription factor Foxo1, known to induce Eomes expression while repressing that of Tbet and inhibiting cell cycle progression \cite{Buchholz2016}, suggesting that the Tbet$^\mathit{lo
}$Eomes$^\mathit{hi
}$ memory precursor cells discussed in Section \ref{SectionSLECMPEC} might adopt a longer cycle than the Tbet$^\mathit{hi
}$ Eomes$^\mathit{lo}$ cells. 

In conclusion, our agent-based multiscale model successfully reproduced several aspects of the CD8 T-cell immune response at both molecular and cellular scales. Even though we cannot infer quantitative conclusions from this study, it suggests that uneven partitioning of molecular content at cell division, as a source of heterogeneity, can modulate cell fate decision and act as a regulator of the magnitude of the response and of the size of the memory cell pool. Actually,  we did not consider intermediaries, namely DNA transcription and mRNA  translation, between gene activation and protein synthesis. Consequently,  our molecular model is an amalgam between gene activity and protein synthesis. Therefore, while our argumentation is based on uneven partitioning of the molecular content, it could also stand for the situation where, when a cell divides,  the two daughter cells inherit different gene activity levels for each gene. All in all, our study focus on molecular heterogeneity generation upon cell division in general, rather than the specific case of molecular partitioning. It stresses that dynamics observed at the cellular scale --- including the initiation of the contraction phase and the origin of memory cells  --- can be explained by structural molecular-content heterogeneity, that is continuously regulated along the response, as CD8 T-cells divide.   
\section*{Conflict of Interest Statement}

The authors declare that the research was conducted in the absence of any commercial or financial relationships that could be construed as a potential conflict of interest.

\section*{Author Contributions}
All co-authors discussed the problem, approach and results. SG, OG and FC designed the model. SG ran simulations and performed analysis. CA and JM conducted the experimental studies. SG  wrote the paper. All the authors approved the final version.  
\section*{Funding}
This work was performed within the framework of the LABEX MILYON (ANR-10-LABX-0070) of Universit\'{e} de Lyon, within the program ``Investissements d'Avenir" (ANR-11-IDEX-0007) operated by the French National Research Agency (ANR) and has been supported by ANR grant PrediVac ANR-12-RPIB-0011.
\section*{Acknowledgments}
 The authors thank the Centre de Calcul de l’Institut National de Physique Nucl\'{e}aire et de Physique des Particules de Lyon (CC-IN2P3) for providing
computing resources, particularly Pascal Calvat and Yonny Cardenas for their valuable help. We also thank the BioSyL Federation and the LabEx Ecofect (ANR-11-LABX-0048) of the University of Lyon for inspiring scientific events. We acknowledge the contribution of SFR Biosciences (UMS3444/CNRS, US8/Inserm, ENS de Lyon, UCBL) facilities. We acknowledge the contributions of the CELPHEDIA Infrastructure (http://www.celphedia.eu/), especially the center AniRA in Lyon.

\section*{Data Availability Statement}
The datasets generated for this study can be found in the Open Science Framework repository \url{https://osf.io/xbq9r}.

mRNA expression data analysed in this study come from the ImmGen project (\url{http://www.immgen.org})

\bibliographystyle{ieeetr}
\bibliography{Bibliographie}

\begin{thebibliography}{10}

\bibitem{Stipdonk}
M.~J. Van~Stipdonk, E.~E. Lemmens, and S.~P. Schoenberger, ``{Na\"{i}ve CTLs
  require a single brief period of antigenic stimulation for clonal expansion
  and differentiation.},'' {\em Nat. Immunol.}, vol.~2, pp.~423--429, May 2001.

\bibitem{Kaech2001}
S.~M. Kaech and R.~Ahmed, ``Memory {CD}8$+$ {T} cell differentiation: initial
  antigen encounter triggers a developmental program in naïve cells,'' {\em
  Nat. Immunol.}, vol.~2, pp.~415--422, may 2001.

\bibitem{Wherry2004}
E.~J. Wherry and R.~Ahmed, ``Memory {CD}8 {T}-cell differentiation during viral
  infection,'' {\em J. Virol.}, vol.~78, no.~11, pp.~5535--5545, 2004.

\bibitem{Badovinac2002}
V.~P. Badovinac, B.~B. Porter, and J.~T. Harty, ``Programmed contraction of
  {CD}8+ {T} cells after infection,'' {\em Nat. Immunol.}, vol.~3,
  pp.~619--626, jun 2002.

\bibitem{Wong2001}
P.~Wong and E.~G. Pamer, ``Cutting edge: Antigen-independent {CD}8 {T} cell
  proliferation,'' {\em J Immunol}, vol.~166, pp.~5864--5868, may 2001.

\bibitem{ANTIA2003}
R.~Antia, C.~T. Bergstrom, S.~S. Pilyugin, S.~M. Kaech, and R.~Ahmed, ``Models
  of {CD}8+ responses: 1. what is the antigen-independent proliferation
  program,'' {\em J. Theor. Biol.}, vol.~221, no.~4, pp.~585--598, 2003.

\bibitem{Blueprint}
B.~E. Russ, A.~E. Denton, L.~Hatton, H.~Croom, M.~R. Olson, and S.~J. Turner,
  ``Defining the molecular blueprint that drives {CD}8+ {T} cell
  differentiation in response to infection,'' {\em Front Immunol}, vol.~3,
  p.~371, 2012.

\bibitem{Blattman2002}
J.~N. Blattman, R.~Antia, D.~J. Sourdive, X.~Wang, S.~M. Kaech,
  K.~Murali-Krishna, J.~D. Altman, and R.~Ahmed, ``Estimating the precursor
  frequency of naive antigen-specific {CD}8 {T} cells,'' {\em J. Exp. Med.},
  vol.~195, no.~5, pp.~657--664, 2002.

\bibitem{Joshi2007}
N.~S. Joshi, W.~Cui, A.~Chandele, H.~K. Lee, D.~R. Urso, J.~Hagman, L.~Gapin,
  and S.~M. Kaech, ``Inflammation directs memory precursor and short-lived
  effector {CD}8+ {T} cell fates via the graded expression of {T}-bet
  transcription factor,'' {\em Immunity}, vol.~27, no.~2, p.~281–295, 2007.

\bibitem{Huang2015}
F.~Huang, W.~Huang, J.~Briggs, T.~Chew, Y.~Bai, S.~Deol, and A.~August, ``The
  tyrosine kinase itk suppresses {CD}8+ memory {T} cell development in response
  to bacterial infection,'' {\em Sci Rep}, vol.~5, p.~7688, 2015.

\bibitem{Lazarevic2013}
V.~Lazarevic, L.~H. Glimcher, and G.~M. Lord, ``T-bet: a bridge between innate
  and adaptive immunity,'' {\em Nat. Rev. Immunol.}, vol.~13, no.~11,
  pp.~777--789, 2013.

\bibitem{Banerjee}
A.~Banerjee, S.~M. Gordon, A.~M. Intlekofer, M.~A. Paley, E.~C. Mooney,
  T.~Lindsten, E.~J. Wherry, and S.~L. Reiner, ``Cutting edge: The
  transcription factor eomesodermin enables {CD}8+ {T} cells to compete for the
  memory cell niche,'' {\em J. Immunol.}, vol.~185, no.~9, pp.~4988--4992,
  2010.

\bibitem{Kaech2012}
S.~M. Kaech and W.~Cui, ``Transcriptional control of effector and memory
  {CD}8$^+$ {T} cell differentiation,'' {\em Nat. Rev. Immunol.}, vol.~12,
  no.~11, pp.~749--761, 2012.

\bibitem{Munitic2010}
I.~Munitic, C.~Evaristo, H.~C. Sung, and B.~Rocha, ``Transcriptional regulation
  during {CD}8 {T}-cell immune responses,'' in {\em Memory T Cells} (M.~Zanetti
  and S.~P. Schoenberger, eds.), pp.~11--27, New York, NY: Springer New York,
  2010.

\bibitem{Joshi2011}
N.~S. Joshi, W.~Cui, C.~X. Dominguez, J.~H. Chen, T.~W. Hand, and S.~M. Kaech,
  ``Increased numbers of preexisting memory {CD}8 {T} cells and decreased
  {T}-bet expression can restrain terminal differentiation of secondary
  effector and memory {CD}8 {T} cells,'' {\em J. Immunol.}, vol.~187, no.~8,
  pp.~4068--4076, 2011.

\bibitem{Arsenio2015}
J.~Arsenio, P.~J. Metz, and J.~T. Chang, ``Asymmetric cell division in {T}
  lymphocyte fate diversification,'' {\em Trends Immunol.}, vol.~36,
  pp.~670--683, nov 2015.

\bibitem{Chang2007}
J.~T. Chang, V.~R. Palanivel, I.~Kinjyo, F.~Schambach, A.~M. Intlekofer,
  A.~Banerjee, S.~A. Longworth, K.~E. Vinup, P.~Mrass, J.~Oliaro, N.~Killeen,
  J.~S. Orange, S.~M. Russell, W.~Weninger, and S.~L. Reiner, ``Asymmetric {T}
  lymphocyte division in the initiation of adaptive immune responses,'' {\em
  Science}, vol.~315, no.~5819, pp.~1687--1691, 2007.

\bibitem{Chang2011}
J.~T. Chang, M.~L. Ciocca, I.~Kinjyo, V.~R. Palanivel, C.~E. McClurkin, C.~S.
  DeJong, E.~C. Mooney, J.~S. Kim, N.~C. Steinel, J.~Oliaro, C.~C. Yin, B.~I.
  Florea, H.~S. Overkleeft, L.~J. Berg, S.~M. Russell, G.~A. Koretzky, M.~S.
  Jordan, and S.~L. Reiner, ``Asymmetric proteasome segregation as a mechanism
  for unequal partitioning of the transcription factor {T}-bet during {T}
  lymphocyte division,'' {\em Immunity}, vol.~34, no.~4, pp.~492--504, 2011.

\bibitem{Ciocca2012}
M.~L. Ciocca, B.~E. Barnett, J.~K. Burkhardt, J.~T. Chang, and S.~L. Reiner,
  ``Cutting edge: Asymmetric memory {T} cell division in response to
  rechallenge,'' {\em J. Immunol.}, vol.~188, pp.~4145--4148, mar 2012.

\bibitem{Cobbold2018}
S.~P. Cobbold, E.~Adams, D.~Howie, and H.~Waldmann, ``{CD}4+ {T} cell fate
  decisions are stochastic, precede cell division, depend on {GITR}
  co-stimulation, and are associated with uropodium development,'' {\em Front
  Immunol}, vol.~9, jun 2018.

\bibitem{Pham2014}
K.~Pham, F.~Sacirbegovic, and S.~M. Russell, ``Polarized cells, polarized
  views: Asymmetric cell division in hematopoietic cells,'' {\em Front
  Immunol}, vol.~5, 2014.

\bibitem{Block1990}
D.~E. Block, P.~D. Eitzman, J.~D. Wangensteen, and F.~Srienc, ``Slit scanning
  of saccharomyces cerevisiae cells: quantification of asymmetric cell division
  and cell cycle progression in asynchronous culture,'' {\em Biotechnol.
  Prog.}, vol.~6, no.~6, p.~504–512, 1990.

\bibitem{Bocharov2013}
G.~Bocharov, T.~Luzyanina, J.~Cupovic, and B.~Ludewig, ``Asymmetry of cell
  division in {CFSE}-based lymphocyte proliferation analysis,'' {\em Front
  Immunol}, vol.~4, p.~264, 2013.

\bibitem{Golding2005}
I.~Golding, J.~Paulsson, S.~M. Zawilski, and E.~C. Cox, ``Real-time kinetics of
  gene activity in individual bacteria,'' {\em Cell}, vol.~123, pp.~1025--1036,
  dec 2005.

\bibitem{Huh2010}
D.~Huh and J.~Paulsson, ``Non-genetic heterogeneity from stochastic
  partitioning at cell division,'' {\em Nat. Genet.}, vol.~43, no.~2,
  p.~95–100, 2010.

\bibitem{Luzyanina2013}
T.~Luzyanina, J.~Cupovic, B.~Ludewig, and G.~Bocharov, ``Mathematical models
  for {CFSE} labelled lymphocyte dynamics: asymmetry and time-lag in
  division,'' {\em J. Math. Biol.}, vol.~69, no.~6-7, p.~1547–1583, 2013.

\bibitem{Sennerstam88}
R.~Sennerstam, ``Partition of protein (mass) to sister cell pairs at mitosis: a
  re-evaluation,'' {\em J. Cell Sci.}, vol.~90(2), no.~11, p.~301–6, 1988.

\bibitem{Thomas2018}
P.~Thomas, G.~Terradot, V.~Danos, and A.~Y. Wei{\ss}e, ``Sources, propagation
  and consequences of stochasticity in cellular growth,'' {\em Nature
  Communications}, vol.~9, oct 2018.

\bibitem{Kinkhabwala2014}
A.~Kinkhabwala, A.~Khmelinskii, and M.~Knop, ``Analytical model for
  macromolecular partitioning during yeast cell division,'' {\em {BMC}
  Biophysics}, vol.~7, sep 2014.

\bibitem{Girel2018}
S.~Girel and F.~Crauste, ``Existence and stability of periodic solutions of an
  impulsive differential equation and application to {CD}8 {T}-cell
  differentiation,'' {\em J. Math. Biol.}, vol.~76, pp.~1765--1795, mar 2018.

\bibitem{Kinjyo2015}
I.~Kinjyo, J.~Qin, S.-Y. Tan, C.~J. Wellard, P.~Mrass, W.~Ritchie, A.~Doi,
  L.~L. Cavanagh, M.~Tomura, A.~Sakaue-Sawano, O.~Kanagawa, A.~Miyawaki, P.~D.
  Hodgkin, and W.~Weninger, ``Real-time tracking of cell cycle progression
  during {CD}8+ effector and memory {T}-cell differentiation,'' {\em Nature
  Communications}, vol.~6, feb 2015.

\bibitem{Yoon2010}
H.~Yoon, T.~S. Kim, and T.~J. Braciale, ``The cell cycle time of {CD}8+ {T}
  cells responding in vivo is controlled by the type of antigenic stimulus,''
  {\em PLoS ONE}, vol.~5, no.~11, p.~e15423, 2010.

\bibitem{Eftimie2016}
R.~Eftimie, J.~J. Gillard, and D.~A. Cantrell, ``Mathematical models for
  immunology: Current state of the art and future research directions,'' {\em
  Bull. Math. Biol.}, vol.~78, pp.~2091--2134, Oct 2016.

\bibitem{Gong2013}
C.~Gong, J.~T. Mattila, M.~Miller, J.~L. Flynn, J.~J. Linderman, and
  D.~Kirschner, ``Predicting lymph node output efficiency using systems
  biology,'' {\em J. Theor. Biol.}, vol.~335, pp.~169--184, 2013.

\bibitem{Gong2014}
C.~Gong, J.~J. Linderman, and D.~Kirschner, ``Harnessing the heterogeneity of
  {T} cell differentiation fate to fine-tune generation of effector and memory
  {T} cells,'' {\em Front Immunol}, vol.~5, 2014.

\bibitem{Prokopiou}
S.~A. Prokopiou, L.~Barbarroux, S.~Bernard, J.~Mafille, Y.~Leverrier, C.~Arpin,
  J.~Marvel, O.~Gandrillon, and F.~Crauste, ``Multiscale modeling of the early
  {CD}8 {T}-cell immune response in lymph nodes: an integrative study,'' {\em
  Computation}, vol.~2, no.~4, pp.~159--181, 2014.

\bibitem{Gao2016}
X.~Gao, C.~Arpin, J.~Marvel, S.~A. Prokopiou, O.~Gandrillon, and F.~Crauste,
  ``{IL}-2~{s}ensitivity and exogenous {IL}-2 concentration gradient tune the
  productive contact duration of {CD}8+ {T} cell-{APC}: a multiscale modeling
  study,'' {\em BMC Syst Biol}, vol.~10, no.~1, p.~77, 2016.

\bibitem{GGH}
F.~Graner and J.~A. Glazier, ``Simulation of biological cell sorting using a
  two-dimensional extended {P}otts model,'' {\em Phys. Rev. Lett.}, vol.~69,
  pp.~2013--2016, 1992.

\bibitem{Jubin2012}
V.~Jubin, E.~Ventre, Y.~Leverrier, S.~Djebali, K.~Mayol, M.~Tomkowiak,
  J.~Mafille, M.~Teixeira, D.~Y.-L. Teoh, B.~Lina, T.~Walzer, C.~Arpin, and
  J.~Marvel, ``T inflammatory memory {CD}8 {T} cells participate to antiviral
  response and generate secondary memory cells with an advantage in {XCL}1
  production,'' {\em Immunol. Res.}, vol.~52, pp.~284--293, apr 2012.

\bibitem{Crauste2017}
F.~Crauste, J.~Mafille, L.~Boucinha, S.~Djebali, O.~Gandrillon, J.~Marvel, and
  C.~Arpin, ``Identification of nascent memory {CD}8 {T} cells and modeling of
  their ontogeny,'' {\em Cell Syst}, vol.~4, pp.~306--317.e4, mar 2017.

\bibitem{Hoyer2008}
K.~K. Hoyer, H.~Dooms, L.~Barron, and A.~K. Abbas, ``Interleukin-2 in the
  development and control of inflammatory disease,'' {\em Immunol. Rev.},
  vol.~226, no.~1, p.~19–28, 2008.

\bibitem{Feinerman2010}
O.~Feinerman, G.~Jentsch, K.~E. Tkach, J.~W. Coward, M.~M. Hathorn, M.~W.
  Sneddon, T.~Emonet, K.~A. Smith, and G.~Altan-Bonnet, ``Single-cell
  quantification of {IL}-2 response by effector and regulatory {T} cells
  reveals critical plasticity in immune response,'' {\em Mol. Syst. Biol.},
  vol.~6, no.~1, p.~437, 2010.

\bibitem{Martins2008}
G.~Martins and K.~Calame, ``Regulation and functions of blimp-1 in {T} and {B}
  lymphocytes,'' {\em Annu. Rev. Immunol.}, vol.~26, no.~1, p.~133–169, 2008.

\bibitem{Yeo2010}
C.~J.~J. Yeo and D.~T. Fearon, ``T-bet-mediated differentiation of the
  activated {CD}8+ {T} cell,'' {\em Eur. J. Immunol.}, vol.~41, no.~1,
  p.~60–66, 2010.

\bibitem{McLane2013}
L.~M. McLane, P.~P. Banerjee, G.~L. Cosma, G.~Makedonas, E.~J. Wherry, J.~S.
  Orange, and M.~R. Betts, ``Differential localization of {T}-bet and eomes in
  {CD}8 {T} cell memory populations,'' {\em J. Immunol.}, vol.~190, no.~7,
  p.~3207–3215, 2013.

\bibitem{Sullivan2003}
B.~M. Sullivan, A.~Juedes, S.~J. Szabo, M.~von Herrath, and L.~H. Glimcher,
  ``Antigen-driven effector {CD}8~{T} cell function regulated by {T}-bet,''
  {\em Proceedings of the National Academy of Sciences}, vol.~100, no.~26,
  p.~15818–15823, 2003.

\bibitem{Bouillet2009}
P.~Bouillet and L.~A. O'Reilly, ``{CD}95, {BIM} and {T} cell homeostasis,''
  {\em Nat. Rev. Immunol.}, vol.~9, no.~7, p.~514–519, 2009.

\bibitem{Strasser2009}
A.~Strasser, P.~J. Jost, and S.~Nagata, ``The many roles of {FAS} receptor
  signaling in the immune system,'' {\em Immunity}, vol.~30, pp.~180--192, feb
  2009.

\bibitem{Kanhere2012}
A.~Kanhere, A.~Hertweck, U.~Bhatia, M.~R. Gökmen, E.~Perucha, I.~Jackson,
  G.~M. Lord, and R.~G. Jenner, ``T-bet and {GATA}3 orchestrate {T}h1 and {T}h2
  differentiation through lineage-specific targeting of distal regulatory
  elements,'' {\em Nat Commun}, vol.~3, p.~1268, 2012.

\bibitem{Shin2009}
H.~Shin, J.~Lee, S.~Park, J.~Chang, and C.~Lee, ``T-bet expression is regulated
  by {EGR}1\mbox{-}{m}ediated signaling in activated {T} cells,'' {\em Clin.
  Immunol.}, vol.~131, no.~3, p.~385–394, 2009.

\bibitem{Suresh}
E.~H. Kim and M.~Suresh, ``Role of {PI3K}/{A}kt signaling in memory {CD}8 {T}
  cell differentiation,'' {\em Front Immunol}, vol.~4, p.~20, 2013.

\bibitem{Cruz}
F.~Cruz-Guilloty, M.~E. Pipkin, I.~M. Djuretic, D.~Levanon, J.~Lotem, M.~G.
  Lichtenheld, Y.~Groner, and a.~Rao, ``Runx3 and {T}-box proteins cooperate to
  establish the transcriptional program of effector {CTL}s,'' {\em J. Exp.
  Med.}, vol.~206, no.~1, pp.~51--59, 2009.

\bibitem{Li2013}
G.~Li, Q.~Yang, Y.~Zhu, H.~Wang, X.~Chen, X.~Zhang, and B.~Lu, ``T-bet and
  eomes regulate the balance between the effector/central memory {T} cells
  versus memory stem like {T} cells,'' {\em {PLoS} {ONE}}, vol.~8, no.~6,
  p.~e67401, 2013.

\bibitem{Ewings2007}
K.~E. Ewings, C.~M. Wiggins, and S.~J. Cook, ``Bim and the pro-survival {B}cl-2
  proteins: Opposites attract, {ERK} repels,'' {\em Cell Cycle}, vol.~6,
  no.~18, p.~2236–2240, 2007.

\bibitem{Kelly2003}
J.~Kelly, R.~Spolski, K.~Imada, J.~Bollenbacher, S.~Lee, and W.~J. Leonard, ``A
  role for {S}tat5 in {CD}8+ {T} cell homeostasis,'' {\em J. Immunol.},
  vol.~170, no.~1, p.~210–217, 2003.

\bibitem{Boyman2010}
O.~Boyman, J.~Cho, and J.~Sprent, ``The role of interleukin-2 in memory {CD8}
  cell differentiation,'' in {\em Memory T Cells} (M.~Zanetti and S.~P.
  Schoenberger, eds.), pp.~28--41, New York, NY: Springer New York, 2010.

\bibitem{last}
J.~D. Ahlers and I.~M. Belyakov, ``Memories that last forever: strategies for
  optimizing vaccine t-cell memory,'' {\em Blood}, vol.~115, no.~9,
  p.~1678–1689, 2009.

\bibitem{Hwang2005}
E.~S. Hwang, ``{IL}-2 production in developing {T}h1 cells is regulated by
  heterodimerization of {RelA} and {T}-bet and requires {T}-bet serine residue
  508,'' {\em J. Exp. Med.}, vol.~202, no.~9, p.~1289–1300, 2005.

\bibitem{Szabo2000}
S.~J. Szabo, S.~T. Kim, G.~L. Costa, X.~Zhang, C.~G. Fathman, and L.~H.
  Glimcher, ``A novel transcription factor, {T}-bet, directs {T}h1 lineage
  commitment,'' {\em Cell}, vol.~100, no.~6, p.~655–669, 2000.

\bibitem{Afkarian2002}
M.~Afkarian, J.~R. Sedy, J.~Yang, N.~G. Jacobson, N.~Cereb, S.~Y. Yang, T.~L.
  Murphy, and K.~M. Murphy, ``T-bet is a {STAT}1-induced regulator of
  {IL}-12{R} expression in naïve {CD}4+ {T} cells,'' {\em Nat. Immunol.},
  vol.~3, no.~6, p.~549–557, 2002.

\bibitem{Baumjohann2013}
D.~Baumjohann and K.~M. Ansel, ``Micro{RNA}-mediated regulation of t helper
  cell differentiation and plasticity,'' {\em Nat. Rev. Immunol.}, vol.~13,
  no.~9, p.~666–678, 2013.

\bibitem{Swat2012}
M.~H. Swat, G.~L. Thomas, J.~M. Belmonte, A.~Shirinifard, D.~Hmeljak, and J.~A.
  Glazier, ``Multi-scale modeling of tissues using {C}ompu{C}ell3{D},'' {\em
  Methods Cell Biol.}, vol.~110, p.~325–366, 2012.

\bibitem{Badovinac2003}
V.~P. Badovinac, K.~A.~N. Messingham, S.~E. Hamilton, and J.~T. Harty,
  ``Regulation of {CD}8+ {T} cells undergoing primary and secondary responses
  to infection in the same host,'' {\em J. Immunol.}, vol.~170, pp.~4933--4942,
  may 2003.

\bibitem{Wirth2010}
T.~C. Wirth, H.-H. Xue, D.~Rai, J.~T. Sabel, T.~Bair, J.~T. Harty, and V.~P.
  Badovinac, ``Repetitive antigen stimulation induces stepwise transcriptome
  diversification but preserves a core signature of memory {CD}8+ {T} cell
  differentiation,'' {\em Immunity}, vol.~33, pp.~128--140, jul 2010.

\bibitem{Gerlach2010}
C.~Gerlach, J.~W. van Heijst, E.~Swart, D.~Sie, N.~Armstrong, R.~M. Kerkhoven,
  D.~Zehn, M.~J. Bevan, K.~Schepers, and T.~N. Schumacher, ``One naive {T}
  cell, multiple fates in {CD}8+ {T} cell differentiation,'' {\em The Journal
  of Experimental Medicine}, vol.~207, pp.~1235--1246, may 2010.

\bibitem{Veiga2000}
H.~Veiga-Fernandes, U.~Walter, C.~Bourgeois, A.~McLean, and B.~Rocha,
  ``{Response of na\"{i}ve and memory CD8+ T cells to antigen stimulation in
  vivo.},'' {\em Nat. Immunol.}, vol.~1, pp.~47--53, July 2000.

\bibitem{Raser2005}
J.~M. Raser, ``Noise in gene expression: Origins, consequences, and control,''
  {\em Science}, vol.~309, pp.~2010--2013, sep 2005.

\bibitem{Lemaitre2013}
F.~Lemaitre, H.~D. Moreau, L.~Vedele, and P.~Bousso, ``Phenotypic {CD}8+ {T}
  cell diversification occurs before, during, and after the first {T} cell
  division,'' {\em J Immunol}, vol.~191, pp.~1578--1585, jul 2013.

\bibitem{Kaech2007}
S.~M. Kaech and E.~J. Wherry, ``Heterogeneity and cell-fate decisions in
  effector and memory {CD}8+ {T} cell differentiation during viral infection,''
  {\em Immunity}, vol.~27, pp.~393--405, sep 2007.

\bibitem{Jung2014}
H.-R. Jung, K.~H. Song, J.~T. Chang, and J.~Doh, ``Geometrically controlled
  asymmetric division of {CD}4+ {T} cells studied by immunological synapse
  arrays,'' {\em {PLoS} {ONE}}, vol.~9, p.~e91926, mar 2014.

\bibitem{Barnett2011}
B.~E. Barnett, M.~L. Ciocca, R.~Goenka, L.~G. Barnett, J.~Wu, T.~M. Laufer,
  J.~K. Burkhardt, M.~P. Cancro, and S.~L. Reiner, ``Asymmetric {B} cell
  division in the germinal center reaction,'' {\em Science}, vol.~335,
  pp.~342--344, dec 2011.

\bibitem{Thaunat2012}
O.~Thaunat, A.~G. Granja, P.~Barral, A.~Filby, B.~Montaner, L.~Collinson,
  N.~Martinez-Martin, N.~E. Harwood, A.~Bruckbauer, and F.~D. Batista,
  ``Asymmetric segregation of polarized antigen on {B} cell division shapes
  presentation capacity,'' {\em Science}, vol.~335, pp.~475--479, jan 2012.

\bibitem{Bevan2001}
M.~J. Bevan and P.~J. Fink, ``The {CD}8 response on autopilot,'' {\em Nat.
  Immunol.}, vol.~2, pp.~381--382, may 2001.

\bibitem{Takemoto2006}
N.~Takemoto, A.~M. Intlekofer, J.~T. Northrup, E.~J. Wherry, and S.~L. Reiner,
  ``Cutting edge: {IL}-12 inversely regulates {T}-bet and eomesodermin
  expression during pathogen-induced {CD}8+ {T} cell differentiation,'' {\em J.
  Immunol.}, vol.~177, no.~11, pp.~7515--7519, 2006.

\bibitem{Buchholz2016}
V.~R. Buchholz, T.~N. Schumacher, and D.~H. Busch, ``T cell fate at the
  single-cell level,'' {\em Annu. Rev. Immunol.}, vol.~34, pp.~65--92, may
  2016.

\end{thebibliography}
\begin{appendix}
\section{Parameter values}
\label{AppendixA}
\begin{center}
{\renewcommand{\arraystretch}{1.1}
\begin{tabular}{|c|l|c|c|}
\hline 
Parameter & Description & Value & Reference  \\ 
\hline 
$A_T$ & target area of CD8 T-cell & 144 $\mu m^2$ & \cite{Gao2016} \\ 
\hline 
$A_{APC}$ & target area of APC & 2250 $\mu m^2$ & calibration \\ 
\hline 
$P_T$ & target perimeter of CD8 T-cell & 48 $\mu m$ & calibration \\ 
\hline 
$\lambda_{pm}$ & weight of perimeter constraint & 10 & calibration \\ 
\hline 
$\lambda_{area}$ & weight of area constraint & 10 & \cite{Prokopiou} \\ 
\hline 
$T$ & temperature & 10 & \cite{Prokopiou} \\ 
\hline 
$J_{APC,m}$ & APC-medium contact energy & 30 & \cite{Prokopiou} \\ 
\hline 
$J_{T,m}$ & CD8 T-cell-medium contact energy & 30 & \cite{Prokopiou} \\ 
\hline 
$J_{APC,APC}$ & APC-APC contact energy & 100 & \cite{Prokopiou}  \\ 
\hline 
$J_{APC,NPA}$ & APC-naive or preactivated CD8 T-cell & 35 & \cite{Prokopiou} \\ 
  &  contact energy &  &  \\ 
\hline 
$J_{APC,AEM}$ & APC-activated, effector or memory  & 100 & calibration  \\ 
&  CD8 T-cell contact energy &  &   \\ \hline 
$J_{T,T}$ & CD8 T-cell-CD8 T-cell contact energy & 100 & \cite{Prokopiou}  \\ 
\hline 
$v(\sigma_{APC})$ & weight of motility energy for an APC & 25 & \cite{Gao2016}  \\ 
\hline 
$v(\sigma_{PA})$ & weight of motility energy for a & 0 & \cite{Gao2016} \\ 
 &  preactivated CD8 T-cell &  &  \\ \hline 
$v(\sigma_{T})$ & weight of motility energy for a & 250 & \cite{Gao2016} \\ 
 &  CD8 T-cell (except preactivated)&  &  \\ \hline 
\end{tabular} 
\captionof{table}{Estimated parameter values for the Cellular Potts model. Weights, temperature and energies are dimensionless parameters.}
\label{paramPotts}}
\end{center}

\begin{center}
{\renewcommand{\arraystretch}{1.1}
\begin{tabular}{|c c c c|}
\hline  
Parameter & Value & Unit & Reference  \\ 
\hline 
Strengths of feedbacks &  &  &    \\ 
\hline 
$\lambda_{R1}$ & 0.58 & mol L$^{-1}$ min$^{-1}$ & calibration  \\ 
\hline 
$\lambda_{R2}$ & 0.007 & min$^{-1}$ & calibration \\ 
\hline 
$\lambda_{E1}$ & 0.001 & min$^{-1}$ & calibration \\ 
\hline 
$\lambda_{T1}$ & 0.07 & mol L$^{-1}$ min$^{-1}$ & calibration  \\ 
\hline 
$\lambda_{T2}$ & 0.06175 & mol L$^{-1}$ min$^{-1}$ & calibration  \\ 
\hline 
$\lambda_{T3}$ & 35 & mol L$^{-1}$ & calibration  \\ 
\hline 
$\lambda_{c1}$ & 0.096 & mol L$^{-1}$ min$^{-1}$ & calibration  \\ 
\hline 
$\lambda_{c2}$ & 0.07 &  L mol$^{-1}$ & calibration  \\ 
\hline 
$\lambda_{c3}$ & 0.23 & / & calibration  \\ 
\hline 
$\lambda_{c4}$ & 1.4 & min$^{-1}$ & calibration  \\ 
\hline 
$\lambda_{E2}$ & 0.073 & L mol$^{-1}$ & calibration  \\ 
\hline 
$\lambda_{E3}$ & 0.06 & mol L$^{-1}$ min$^{-1}$ & calibration  \\ 
\hline 
$\lambda_{E4}$ & 0.09 & mol L$^{-1}$ min$^{-1}$ & calibration  \\ 
\hline 
$\lambda_{E5}$ & 20 & / & calibration  \\ 
\hline 
$\lambda_{E6}$ & 10 & mol L$^{-1}$ & calibration  \\ 
\hline 
$\lambda_{E7}$ & 0.035 & L mol$^{-1}$ & calibration  \\ 
\hline
Degradation rates &  &  & \\ 
\hline 
$k_R$ & 0.0077 & min$^{-1}$ & calibration  \\ 
\hline 
$k_e$ & 0.0154 & min$^{-1}$ & calibration  \\ 
\hline 
$k_T$ & 0.00051 & min$^{-1}$ & calibration \\ 
\hline 
$k_{F}$ & 0.003 & min$^{-1}$ & calibration  \\ 
\hline 
$k_c$ & 0.0038 & min$^{-1}$ & \cite{Prokopiou}\\ 
\hline 
$k_E$ & 0.0035 & min$^{-1}$ & calibration \\ 
\hline 
Association/dissociation rates &   &   & calibration \\ 
\hline 
$\mu_{IL2}^+$ & $6\times 10^6$ & L mol$^{-1}$ min$^{-1}$ & calibration \\ 
\hline 
$\mu_{IL2}^-$ & 0.12 & min$^{-1}$ & calibration \\ 
\hline 
$\mu_F^+$ & 0.0002&  L mol$^{-1}$ min$^{-1}$ & \cite{Gao2016} \\ 
\hline 
$\mu_F^-$ & 0.004 & min$^{-1}$ & \cite{Prokopiou} \\ 
\hline 
$k_E$ & 0.0035 & min$^{-1}$ & calibration \\ 
\hline 
Other &  &  &  \\ 
\hline 
$\lambda_F$ & 4.2$\times 10^{-5}$ & mol L$^{-1}$ min$^{-1}$ & \cite{Prokopiou} \\ 
\hline 
$n$ & 3 & / & calibration \\ 
\hline 
\end{tabular}
\captionof{table}{Parameter values for System (1)-(6). }
\label{paramIntra}}
\end{center}

{\renewcommand{\arraystretch}{1.1}
\begin{center}
\begin{tabular}{|c c c c|}
\hline 
Parameter & Value & Unit & Reference  \\ 
\hline 
$\lambda_{R3}$ & 5$\times 10^{-9}$  &  mol L$^{-1}$ min$^{-1}$ & calibration  \\ 
\hline 
$\lambda_{R4}$ & 15 & mol L$^{-1}$  & calibration  \\ 
\hline 
$\lambda_{T4}$ & 0.08 & L mol$^{-1}$  & calibration  \\ 
\hline 
$\lambda_1$ & $3\times 10^{-8}$ & L mol$^{-1}$  & calibration  \\ 
\hline 
$D$ & 1.776 &  $\mu m^2$ min$^{-1}$ & calibration  \\ 
\hline 
$\delta$ & 0.187 & min$^{-1}$ & \cite{Gao2016}  \\ 
\hline 
\end{tabular} 
\captionof{table}{Parameters values for equation (7).}
\label{paramIL2}
\end{center}}

\begin{center}
{\renewcommand{\arraystretch}{1.1}
\begin{tabular}{|c c c c|}
\hline  
Thresholds & Value & Unit & \\ 
\hline 
$IL2R_{th}$ & 62.9 & mol L$^{-1}$  & \\ 
\hline 
$Tbet_{th}$ & 16.8 & mol L$^{-1}$ & \\ 
\hline 
$Eomes_{th}$ & 16 (15.6 in Section 3.4) & mol L$^{-1}$  &\\ 
\hline 
$Caspases_{th}$ & 19.42 & mol L$^{-1}$ & \\ 
\hline 
\end{tabular}
\captionof{table}{Threshold values for differentiation and death of CD8 T-cells.}
\label{paramThresholds} }
\end{center}

\section{Sensitivity analysis to parameter $Eomes_{th}$}
\label{AppendixB}
 In this paper, we enriched previous models \cite{Gao2016,Prokopiou} in order to allow differentiation into memory cell.  In our model, this differentiation occurs when the concentration of protein Eomes in an activated or effector cell crosses the threshold $Eomes_{th}$ (see Section 2.3.1). Here, we discuss how sensitive our model is to this parameter. Sensitivity to other parameters has been investigated in \cite{Gao2016} and \cite{Prokopiou}.

\begin{figure}[tp!]
\centering
\includegraphics[width=10cm]{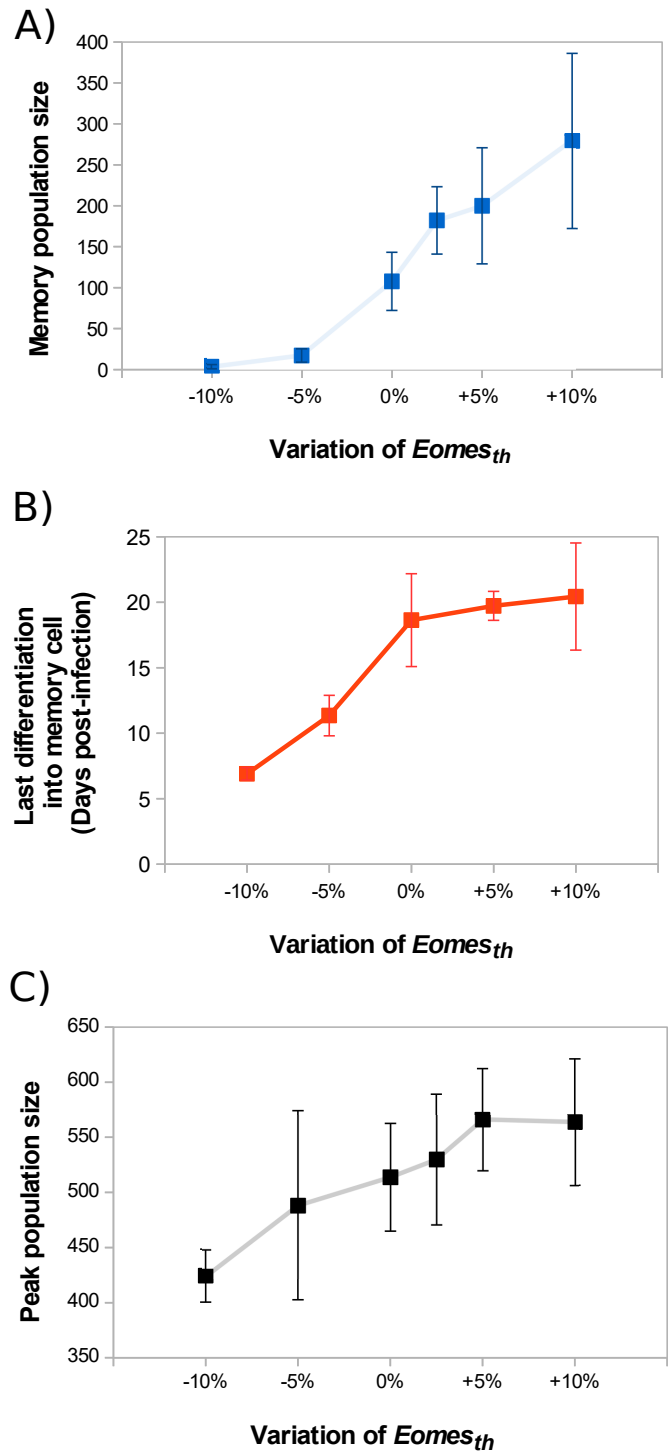}
   \caption{Sensitivity analysis to parameter $Eomes_{th}$. (A) Number of memory cells at the end of the response D25 p.i., (B) day at which the last differentiation into memory cell occurs and (C) number of CD8 T-cells at the peak of the response as functions of the parameter $Eomes_{th}$ (mean +/- standard deviation over 5 simulations). Results are shown for variations of $-10\%$, $-5\%$, $+2.5\%$, $+5\%$ and $+10\%$ from the base value of $Eomes_{th}$ (Table \ref{paramThresholds}).} 
   \label{FigureASEomes}
\end{figure}

 It is no surprise that the number of memory cells at the end of the response (D25 p.i) is highly sensitive to $Eomes_{th}$ value, as shown on Figure \ref{FigureASEomes}.A. 	Indeed, for low values of $Eomes_{th}$, the memory precursor cells identified in Section 3.3 differentiate early   while for higher values of $Eomes_{th}$ they can accomplish a few additional rounds of division before their Eomes concentrations are sufficiently high to trigger memory differentiation, leading to bigger memory cell populations. Indeed, Figure \ref{FigureASEomes}.B illustrates that the lower the value of $Eomes_{th}$, the sooner the pool of memory CD8 T-cells stops to expand.

 We can see on Figure \ref{FigureASEomes}.C that the number of cells at the peak of the response (i.e. the maximal expansion size reached by the CD8 T-cell population) is less sensitive to the value $Eomes_{th}$. However, it appears that the peak population size slightly increases with the value of $Eomes_{th}$. This is mainly due to the reason exposed above, i.e. memory precursor effector cells proliferate more before their differentiation into non-dividing memory cells.  

It is also worth noting that if $Eomes_{th}$ is too low, CD8 T-cells can differentiate into memory cells even if their molecular content does not match with the expected properties of memory cells (survival and low cytotoxicity). Indeed, if $Eomes_{th}$ is $90\%$ (resp. $95\%$) of its base value (see Table \ref{paramThresholds}),  an average of  3.6 (resp. 0.4) memory cells die during the response (mean over 5 simulations, results not shown) while for the base value of $Eomes_{th}$ or upper values there is no memory cell death.

\begin{figure}[tp!]
\centering
\includegraphics[width=10cm]{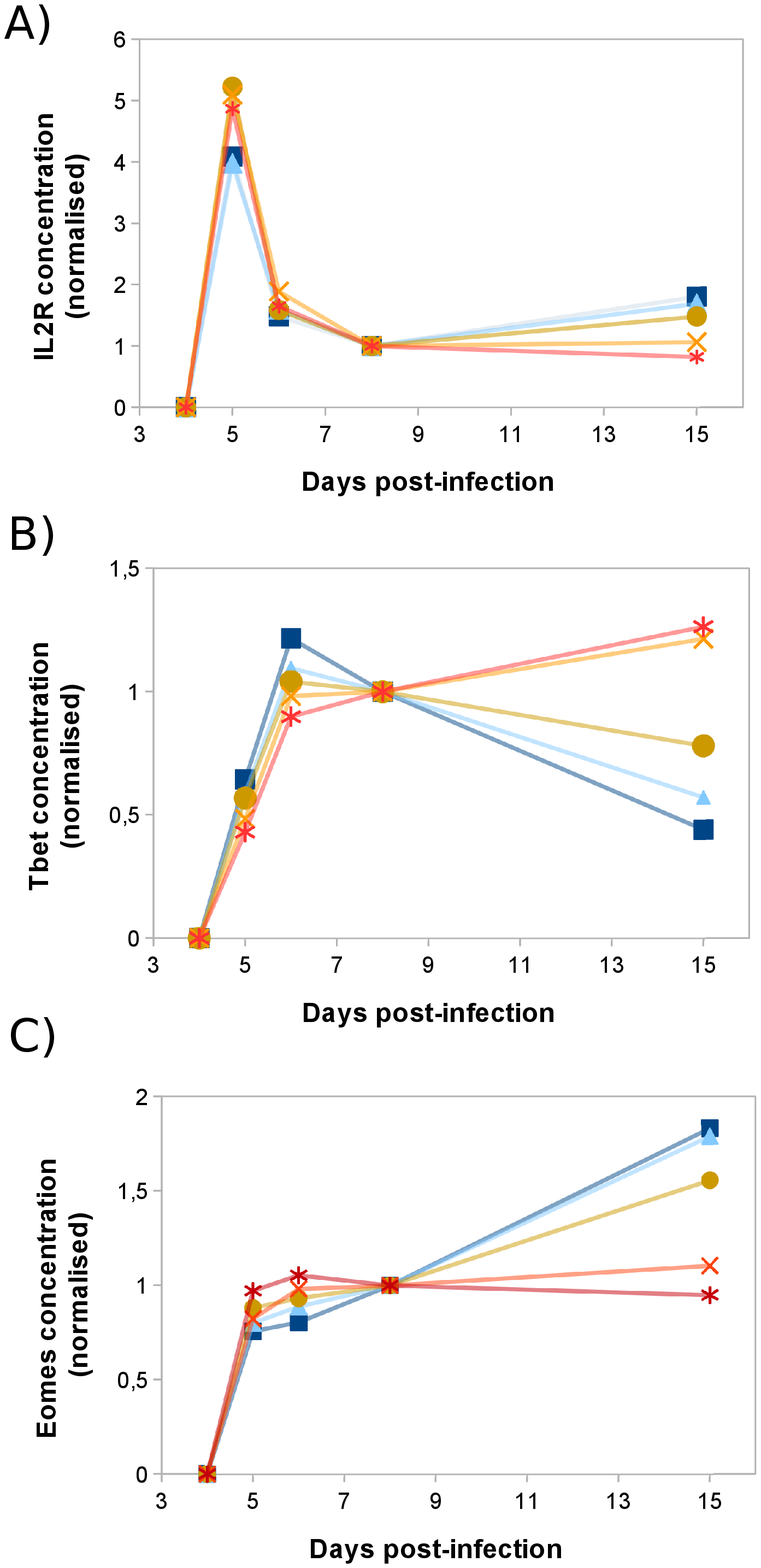}
   \caption{Mean concentration of (A) IL2 receptors, (B) Tbet and (C) Eomes among the CD8 T-cell population normalised by the concentration value D8 p.i. for different values of $Eomes_{th}$ (mean over 5 simulations).  Results are shown for variations of $-10\%$ (red stars), $-5\%$ (orange crosses), $0\%$ (yellow discs), $+5\%$ (light blue triangles) and $+10\%$ (dark blue squares) from the base value of $Eomes_{th}$ (Table \ref{paramThresholds}). } 
   \label{FigureASEomesMolec}
\end{figure}

Figure \ref{FigureASEomesMolec} represents the mean concentration of IL2 receptors, Tbet and Eomes  among the CD8 T-cell population from D4 to D15p.i., for different values of $Eomes_{th}$. Since Eomes expression is associated with memory phenotype, it is no surprise that Eomes concentration increases more slowly, or even decreases, for low values of $Eomes_{th}$. On the opposite, high values of $Eomes_{th}$,  associated with big memory cell populations, lead to decreasing Tbet concentrations.

\end{appendix}

\end{document}